\documentstyle[aaspptwo,psfig]{article}

\begin{document}
\title{The circumstellar shell of the 
post-AGB star HD~56126: the $^{12}$CN/$^{13}$CN isotope ratio and
fractionation}
\author{Eric J. Bakker  and David L. Lambert}
\affil{Department of Astronomy and  W. J. McDonald Observatory, 
University of Texas, Austin, TX 78712-1083, USA,
ebakker@astro.as.utexas.edu, dll@astro.as.utexas.edu}

\begin{abstract}
We have detected circumstellar absorption
lines of the $^{12}$CN and $^{13}$CN  Violet 
and Red System in the spectrum of 
the post-AGB star HD~56126. 
From a synthetic spectrum analysis, we 
derive a Doppler broadening parameter
of $b=0.51\pm0.04$ km~s$^{-1}$,
$^{12}$CN/$^{13}$CN$=38\pm2$, and a lower limit
of $2000$ on $^{12}$CN/$^{14}$CN and 
$^{12}$C$^{14}$N/$^{12}$C$^{15}$N.
A simple chemical model has been computed of the circumstellar shell
surrounding HD~56126 that takes into account the gas-phase ion-molecule
reaction between CN and C$^{+}$. From this we infer that
this reaction leads to isotopic fractionation of CN.
Taking into account the isotopic exchange reaction
and the observed $^{12}$CN/$^{13}$CN 
we find $^{12}$C/$^{13}$C$\sim 67$ (for $T_{\rm kin}=25$ K).
Our analysis suggests that $^{12}$CN has a somewhat higher
rotational temperature than $^{13}$CN: $T_{\rm rot}=11.5\pm0.6$
and $8.0\pm0.6$ K respectively. We identify possible causes
for this difference in excitation temperature, among which
the $N''$ dependence of the isotopic exchange reaction.

\noindent
\end{abstract}

\section{Introduction}

The evolution of a low-mass red giant terminates 
on the tip of the Asymptotic Giant Branch (AGB).
The remnant, which is a carbon-oxygen core with a dilute extended
convective envelope, evolves along a constant luminosity
track toward the White Dwarf (WD) phase. This transition from
the AGB to the WD is referred to as the post-AGB phase, or
the Pre-Planetary Nebulae (PPN) phase.
Since low-mass stars are the main contributors of carbon, nitrogen,
and s-process elements to the interstellar medium (ISM)
(Forestini \& Charbonnel 1997), accurate information 
of the chemical composition of AGB and post-AGB stars is 
important.
Post-AGB stars may provide useful information on the composition
of the gas returned to the ISM. In particular, the fact that their
spectra are much simpler than those of the cool AGB stars affords
novel opportunities to infer details of the composition
of AGB stellar envelopes.

The present photosphere of HD~56126 ($T_{\rm eff} = 7000$ K)
is carbon-rich (C/O$\simeq$1.4),
metal-poor ([Fe/H]$\simeq$-1.0), and enhanced 
in s-process elements ([s/Fe]=1.7) (Klochkova 1995).
Such an abundance  pattern resembles that of carbon stars
(Lambert et al. 1986, Utsumi 1970) 
that result from the dredge-up of nucleosynthesis products
as a star evolves up the AGB.
HD~56126 is the prototype of a group of post-AGB and AGB
stars that exhibit absorption lines from circumstellar 
C$_{2}$ and CN (from now on C means $^{12}$C and N means $^{15}$N)
in their spectra (Bakker et al. 1996 (Paper I),
Bakker et al. 1997 (Paper~II)). In order to facilitate
further discussion, we introduce the term
``C2CN'' stars to refer to this group of stars. 
Molecules are present in the detached shell surrounding
the star and the chemical composition of the dusty shell
reflects the photospheric composition of the star when it
was at the tip of the  Asymptotic Giant Branch (TP-AGB).
The molecular composition of the shell has evolved, for
example a simple molecule like CN is believed to be formed as a result
of photodissociation of complex molecules (in this case HCN)
by the interstellar radiation field.

Excitation of the CN molecule's X$^{2}\Sigma^{+}$ ground state
is likely controlled by absorption and re-emission of photons 
in the Violet (${\rm B}^{2}\Sigma^{+} \rightarrow {\rm X}^{2}\Sigma^{+}$)
and Red  (${\rm A}^{2}\Pi \rightarrow {\rm X}^{2}\Sigma^{+}$) System,
and pure rotational transitions (principally de-excitation) in
the X$^{2}\Sigma^{+}$ state. At the low densities of the circumstellar
shell, the pure rotational transitions will cool the rotational ladder
below the gas kinetic temperature
(sub-thermal). For a symmetric molecule without a dipole moment
(e.g. C$_{2}$), pure rotational transitions are forbidden and 
the molecule can not efficiently cool. The excitation temperature
is higher than the kinetic temperature of the gas (supra-thermal).
Through measurement of the 
level population and modeling of the (de-)excitation
processes, the CN molecule serves as a probe of the physical
conditions in the shell. A molecule like CN additionally affords
an opportunity to compare column densities of isotopomers, e.g.
the ratio of CN to $^{13}$CN. In turn, understanding of the chemistry
enables the isotopic ratio $^{12}$C/$^{13}$C to be derived from the
measured $^{12}$CN/$^{13}$CN ratio.

In this paper (Paper~III)
we present and discuss the first measurement of the
CN/$^{13}$CN ratio and an estimate of the 
$^{12}$C/$^{13}$C ratio and lower limits on
$^{12}$C/$^{14}$C and $^{14}$N/$^{15}$N in the circumstellar shell of a
post-AGB star.
In Sec. 2 we discuss the observations and the dataset of equivalent
widths used in our analysis. Sec.~3 goes into the details of
the molecular parameters used,
and Sec.~4 describes the analysis and results. A discussion
of the results is presented in Sec.~5.

\section{Observations and equivalent widths}

High-resolution spectra
were obtained with 2.7~m Harlan J. Smith telescope
of the  W. J. McDonald observatory and the coud\'{e} cross-dispersed
echelle spectrograph
(Tull et al. 1995). Spectra at a resolution $R = \lambda/\Delta\lambda
\simeq 140,000$ were acquired of the ($v'-v'')=(0-0)$ 
band of the CN Violet System and $R \simeq 200,000$
of the (3-0) and (4-0) bands of the CN Red System.  Observations were made
from  December 1996 to March 1997 (Table~1).
The spectral resolution of the spectra was determined from the
$FWHM$ of the emission lines in the accompanying ThAr arc spectrum 
and corrected for the intrinsic width of the ThAr emission lines.

\begin{table*} % Table~1
\caption{Log of observations.}
\centerline{\begin{tabular}{lllllrl} % 7 columns
          &            &             &       &               &   &      \\
\hline
\hline
Date      &$HJD$       &Int. [s]     &$R=\frac{\lambda}{\Delta \lambda}$
&$\lambda_{\rm c}$ [\AA]&$SNR$&Remark \\
\hline
          &            &             &       &               &   &       \\
29 Dec. 96&2450446.7158&6$\times$1800&200,000&7558           &100&Red (3,0) + H$\alpha$        \\
02 Jan. 97&2450450.8708&3$\times$1800&140,000&4062           & 20&Violet (0,0) + Ca~II~H + H$\epsilon$\\
28 Feb. 97&2450507.5840&6$\times$1800&140,000&4056           & 15&Violet (0,0) + H$\delta$        \\
01 Mar. 97&2450508.5696&6$\times$1800&140,000&3874           & 22&Violet (0,0) + H$\delta$        \\
03 Mar. 97&2450510.6087&8$\times$1800&200,000&6086           & 70&Red (3,0) + Red (4,0)            \\
\multicolumn{7}{l}{Co-added spectra}                               \\
\multicolumn{2}{l}{Violet System (0,0)}&7.5 hrs &140,000&3877& 35& \\
\multicolumn{2}{l}{Red    System (3,0)}&7.0 hrs &200,000&6945&200& \\
\multicolumn{2}{l}{Red    System (4,0)}&4.0 hrs &200,000&6195& 90& \\
\hline
\hline
\end{tabular}}
\end{table*}

Observations of 30 minutes have been co-added to give the final
spectra for each night. 
All spectra were reduced using IRAF in the standard manner.
Individual integrations were  corrected for
the velocity shift due to the Earth rotation and instrumental drift
 before combining
them to an average spectrum. 
The final S/N ratio of the spectra is 35 for the Violet (0-0) band 
and 200 and 90 for the Red (3-0)  and (4-0) bands respectively.

\begin{figure*} % Fig.~1
\centerline{\hbox{\psfig{figure=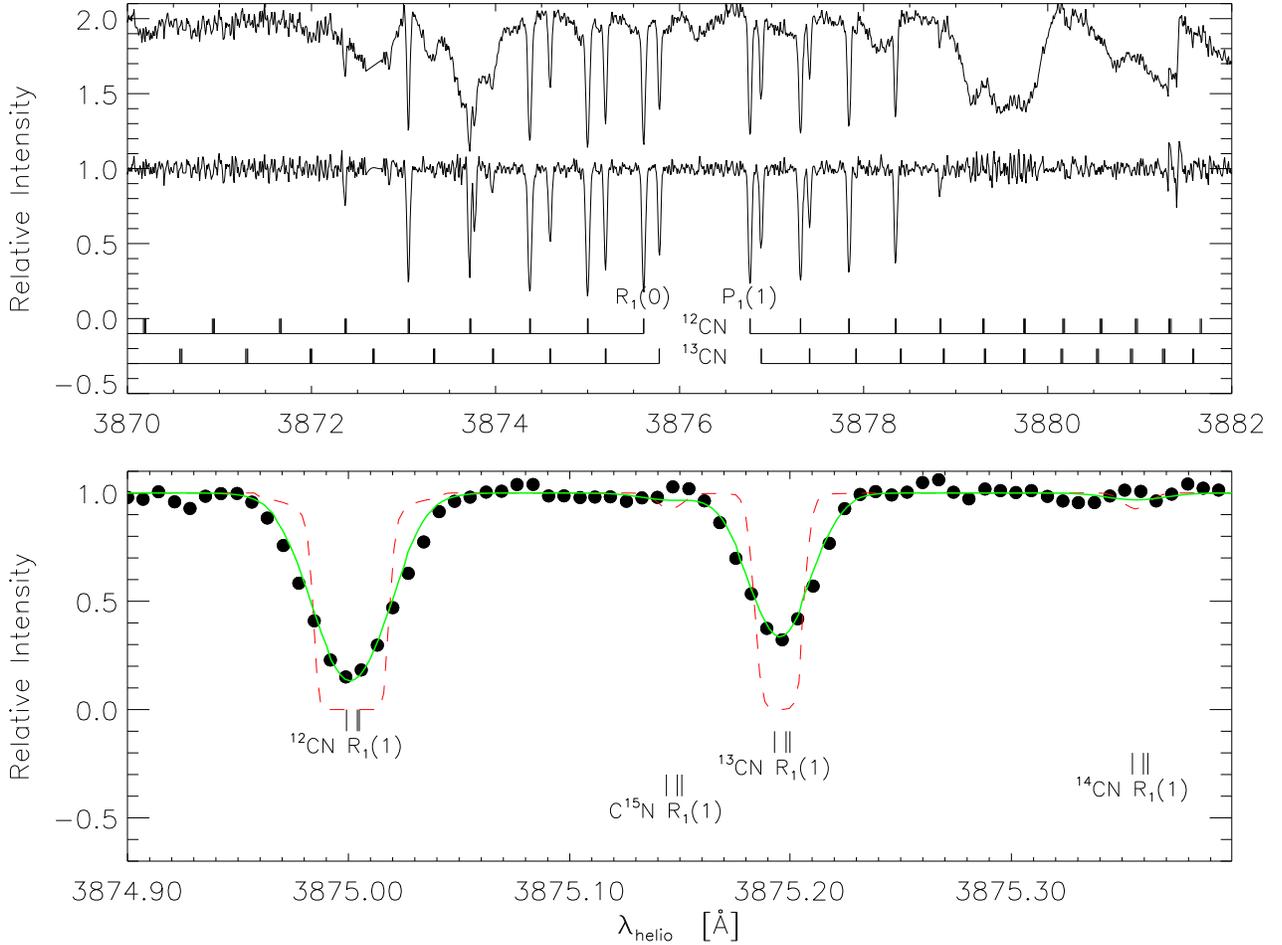,width=\textwidth}}}
\caption{CN Violet System (0,0) band towards HD~56126. 
The upper panel shows the observed spectrum and
a  rectified spectrum corrected
for the underlying photospheric features.
The rectified spectrum contains only circumstellar lines.
The lower panel shows on an expanded wavelength scale
the strongest line (the R$_{1}$, R$_{2}$, and $^{\rm R}Q_{21}$ blend 
for $N''=1$),
and demonstrates that we have not detected $^{14}$CN and C$^{15}$N.
The dashed spectrum is a synthetic spectrum computed for $b=0.51$ km~s$^{-1}$~,
$T_{\rm rot}=11.5$ K for CN, $T_{\rm rot}=8.0$  for the CN isotopes,
and the isotope or lower limit isotope ratios as determined
in this work. The solid spectrum is the synthetic spectrum 
convolved to our spectral resolution of $R=140,000$.}
\end{figure*}

\begin{figure*} % Fig.~2
\centerline{\hbox{\psfig{figure=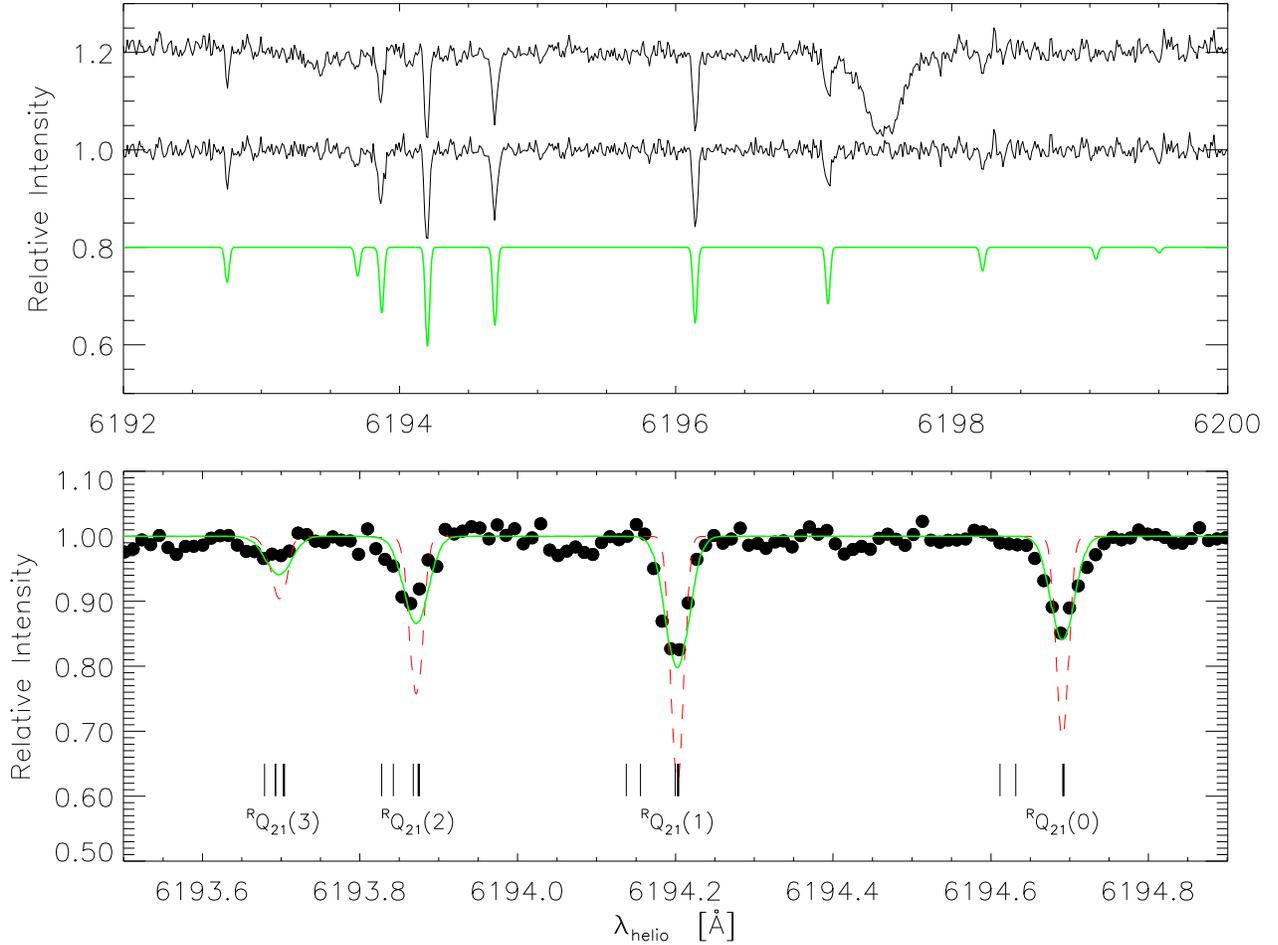,width=\textwidth}}}
\caption{CN Red System (4,0) band towards HD~56126. 
The upper panel shows the observed spectrum (offset by 0.2),
a  rectified spectrum corrected 
for the underlying photospheric features, and convolved 
synthetic spectrum (offset by -0.2)
The rectified spectrum contains only circumstellar lines.
The lower panel shows 
on an expanded wavelength scale centered around
the strongest line (the $\rm^{R}Q_{21}$(1) or R$_{2}$(1) blend) 
(lower panel).
The dashed and solid synthetic spectrum are explained in the caption
of Fig.~1. The $^{13}$CN (4,0) band is red-shifted  by about 
54 \AA~ and not covered by this spectrum. 
The solid spectrum is the synthetic spectrum 
convolved to our spectral resolution of $R=200,000$.}
\end{figure*}

Portion of the final spectra are shown in Figs.~1 and 2.
From each final spectrum we removed the molecular features and fitted
a high order spline that followed all
photospheric features. The observed spectrum what divided by this
fit to obtain the rectified spectrum which only contains the molecular
features. Circumstellar
CN lines are readily distinguishable 
from the much broader photospheric
lines of various atomic species.
The equivalent widths (Tables~2 and 3) of the molecular lines were measured
relative to the local continuum (taking into account stellar
absorption lines). A search for $^{14}$CN and C$^{15}$N 
of the Violet System (0,0) band 
was unsuccessful: an upper limit to their equivalent width 
is 1.5 m\AA.

HD~56126 is a pulsating star. The photospheric spectrum from the
star might therefore change from one run to the other and combining
the spectrum from different runs could lead to errors.
We therefore have measured the equivalent width of the lines for
each night separately and averaged to obtain the final equivalent
width. The spectra in Figs.~1 and 2 are combined using all available
data and there is no indication that 
the photospheric spectrum has indeed changed from one run to the next.

This high quality dataset was extended with previous data of
the Red System (1,0), (2,0), and (3,0) bands (presented in Paper~I),
provided by spectra collected with the WHT/UES at a spectral
resolution of $R\approx 50,000$.
Our new measurements for the Red (3-0) band are in good
agreement with measurements of Paper I.
The weaker lines were only detected at
our higher resolution. 
Our new measurements for the Red (3-0) band are in good
agreement with measurements of Paper I.

\begin{table*} % Table~2
\caption{CN Violet System (0,0) with $f_{(0,0)}=0.0330$. 
Derived column densities are given in Table~4.}
\centerline{\begin{tabular}{rllllllrrl}
\hline
\hline
$B(N'')$&$\lambda_{\rm rest}$  [\AA]&
\multicolumn{3}{c}{$\Delta \lambda$  [\AA]$^{a}$}&$f_{N'J',N''J''}$
&$f_{\rm eff}N$[cm$^{-2}$] &
\multicolumn{2}{c}{$W_{\lambda}$ [m\AA]}&Remark \\
\cline{3-5} 
\cline{8-9}
&CN&$^{13}$CN&$^{14}$CN&C$^{15}$N&
& &CN&$^{13}$CN& \\
\hline
$\rm  R_{1 }$(6)&3870.666l&0.348l&0.604c&0.239c&0.0176&$0.0178N(6)$&$<2.0$      &$<2.0$      &\\
$\rm  R_{2 }$(6)&3870.657l&      &      &      &0.0178&            &            &            &\\
$\rm^RQ_{21}$(6)&3870.665c&      &      &      &0.0002&            &            &            &\\
\cline{1-6}
$\rm  R_{1 }$(5)&3871.358l&0.300l&0.550c&0.218c&0.0178&$0.0180N(5)$&$ 9.6\pm0.3$&$<2.0$      &\\
$\rm  R_{2 }$(5)&3871.366l&      &      &      &0.0180&            &            &            &\\
$\rm^RQ_{21}$(5)&3871.371c&      &      &      &0.0002&            &            &            &\\
\cline{1-6}
$\rm  R_{1 }$(4)&3872.045l&0.277l&0.497c&0.197c&0.0180&$0.0183N(4)$&$29.0\pm0.5$&$<1.7$      &too strong\\
$\rm  R_{2 }$(4)&3872.053l&      &      &      &0.0183&            &            &            &\\
$\rm^RQ_{21}$(4)&3872.058c&      &      &      &0.0003&            &            &            &\\
\cline{1-6}
$\rm  R_{1 }$(3)&3872.716l&0.252l&0.446c&0.176c&0.0183&$0.0189N(3)$&$25.1\pm0.3$&$ 6.3\pm0.5$&perturbed \\
$\rm  R_{2 }$(3)&3872.712l&      &      &      &0.0189&            &            &            &\\
$\rm^RQ_{21}$(3)&3872.725c&      &      &      &0.0006&            &            &            &\\
\cline{1-6}
                &3872.774l&      &      &      &      &            &$10.0\pm0.3$&            &perturbation\\
\cline{1-6}
$\rm  R_{1 }$(2)&3873.363l&0.222l&0.396c&0.156c&0.0189&$0.0198N(2)$&$37.1\pm0.3$&$18.6\pm0.3$&\\
$\rm  R_{2 }$(2)&3873.370l&      &      &      &0.0198&            &            &            &\\
$\rm^RQ_{21}$(2)&3873.371l&      &      &      &0.0009&            &            &            &\\
\cline{1-6}
$\rm  R_{1 }$(1)&3873.991l&0.192l&0.348c&0.137c&0.0198&$0.0220N(1)$&$36.7\pm0.3$&$25.0\pm0.3$&\\
$\rm  R_{2 }$(1)&3873.996l&      &      &      &0.0220&            &            &            &\\
$\rm^RQ_{21}$(1)&3873.998l&      &      &      &0.0022&            &            &            &\\
\cline{1-6}
$\rm  R_{1 }$(0)&3874.602l&0.176l&0.301c&0.119c&0.0220&$0.0330N(0)$&$36.3\pm0.3$&$21.1\pm0.4$&\\
$\rm^RQ_{21}$(0)&3874.605l&      &      &      &0.0110&            &            &            &\\
\cline{1-6}
$\rm^PQ_{12}$(1)&3875.758l&      &      &      &0.0110&            &            &            &\\
$\rm  P_{1 }$(1)&3875.759l&0.130o&0.212c&0.083c&0.0110&$0.0110N(1)$&$32.6\pm0.3$&$18.4\pm0.3$&\\
\cline{1-6}
$\rm^PQ_{12}$(2)&3876.304l&      &      &      &0.0022&            &            &            &\\
$\rm  P_{1 }$(2)&3876.306l&0.105l&0.170c&0.067c&0.0132&$0.0132N(2)$&$29.0\pm0.3$&$13.7\pm0.3$&\\
$\rm  P_{2 }$(2)&3876.308l&      &      &      &0.0110&            &            &            &\\
\cline{1-6}
$\rm^PQ_{12}$(3)&3876.830c&      &      &      &0.0009&            &            &            &\\
$\rm  P_{1 }$(3)&3876.834l&0.063l&0.130c&0.050c&0.0141&$0.0141N(3)$&$27.6\pm0.3$&$ 4.2\pm0.5$&\\
$\rm  P_{2 }$(3)&3876.836l&      &      &      &0.0132&            &            &            &\\
\cline{1-6}
$\rm^PQ_{12}$(4)&3877.336c&      &      &      &0.0006&            &            &            &\\
$\rm  P_{1 }$(4)&3877.341l&0.056l&0.091c&0.035c&0.0147&$0.0147N(4)$&$24.1\pm0.3$&$<1.0$      &\\
$\rm  P_{2 }$(4)&3877.345l&      &      &      &0.0141&            &            &            &\\
\cline{1-6}
$\rm^PQ_{12}$(5)&3877.821c&      &      &      &0.0003&            &            &            &\\
$\rm  P_{1 }$(5)&3877.828l&0.028c&0.054c&0.020c&0.0150&$0.0150N(5)$&$ 3.9\pm0.5$&polluted    &perturbed\\
$\rm  P_{2 }$(5)&3877.825l&      &      &      &0.0147&            &            &            &\\
\cline{1-6}
                &3877.885l&      &      &      &      &            &$<1.0$      &            &perturbation\\
\cline{1-6} 
$\rm^PQ_{12}$(6)&3878.286c&      &      &      &0.0002&            &            &            &\\
$\rm  P_{1 }$(6)&3878.294l&0.008c&0.018c&0.006c&0.0152&$0.0152N(6)$&$<1.0$      &$<1.0$      &\\
$\rm  P_{2 }$(6)&3878.300l&      &      &      &0.0150&            &            &            &\\
\hline
\hline
\multicolumn{10}{l}{a: isotopic shift relative to CN. 
 Example: R$_{2}$(1) $^{14}$CN at 3873.996+0.348=3874.344 \AA} \\
\multicolumn{10}{l}{c: computed; l: laboratory data; o: spectrum} \\
\multicolumn{10}{l}{bl: blended with the $^{12}$CN P(5) blend} \\
\multicolumn{10}{l}{For lines with laboratory data: $v$(CN)$=77.5\pm0.5$ km~s$^{-1}$
                    and $v$($^{13}$CN)$=76.7\pm1.8$ km~s$^{-1}$} \\
\end{tabular}}
\end{table*}

\begin{table*} % Table~3
\caption{CN Red System (4,0) with $f_{(4,0)}=1.09\times 10^{-4}$.
Derived column densities are given in Table~4.}
\centerline{\begin{tabular}{rlllll}
\hline
\hline
$B(N'')$        &$\lambda_{\rm rest}$ [\AA]
&$f_{N'J',N''J''}\times10^{4}$ 
&$f_{\rm eff}\times 10^{4}N(N'')$[cm$^{-2}$]&$W_{\lambda}$ [m\AA]&
Remark \\ 
\hline
$\rm ^{S}R_{21}$(3)&6186.293l&0.101&0.058N(3)&$\leq1.5$  & \\
\cline{1-4}
$\rm ^{S}R_{21}$(2)&6187.764l&0.112&0.067N(2)&$1.8\pm0.5$& \\
\cline{1-4}
$\rm ^{S}R_{21}$(1)&6189.397l&0.127&0.085N(1)&$3.1\pm0.5$& \\
\cline{1-4}
$\rm ^{S}R_{21}$(0)&6191.159c&0.151&0.151N(0)&$2.4\pm0.1$& \\
\cline{1-4}
$\rm ^{ }R_{2 }$(3)&6192.087l&0.202&         &           & \\
$\rm ^{R}Q_{21}$(3)&6192.101c&0.217&0.211N(3)&$\leq 1.5$ & \\
\cline{1-4}
$\rm ^{ }R_{2 }$(2)&6192.255l&0.203&         &           & \\
$\rm ^{R}Q_{21}$(2)&6192.274c&0.237&0.223N(2)&$5.3\pm0.6$& \\
\cline{1-4}
$\rm ^{ }R_{2 }$(1)&6192.572c&0.215&         &           & \\
$\rm ^{R}Q_{21}$(1)&6192.605l&0.264&0.248N(1)&$8.0\pm0.2$& \\
\cline{1-4}
$\rm ^{R}Q_{21}$(0)&6193.008l&0.363&0.363N(0)&$7.2\pm0.1$& \\
\cline{1-4}
$\rm ^{ }Q_{2 }$(1)&6194.485l&0.363&0.182N(1)&$6.7\pm0.1$& \\
$\rm ^{Q}P_{21}$(1)&6194.550c&0.091&         &           & \\
\cline{1-4}
$\rm ^{ }Q_{2 }$(2)&6195.467l&0.316&0.188N(2)&$4.5\pm0.5$& \\
$\rm ^{Q}P_{21}$(2)&6195.515c&0.102&         &           & \\
\cline{1-4}
$\rm ^{ }Q_{2 }$(3)&6196.622l&0.323&0.197N(3)&$1.7\pm0.2$& \\
$\rm ^{Q}P_{21}$(3)&6196.638c&0.103&         &           & \\
\cline{1-4}
$\rm ^{ }P_{2 }$(2)&6197.458l&0.091&0.036N(2)&$\leq 1.5$ & \\
\hline
\hline
\end{tabular}}
\end{table*}

\section{Molecular data}
\subsection{CN Violet System}

Wavelengths for CN Violet System (0,0) were computed from the
wavenumbers of Prasad et al. (1992) and the isotopic
wavelengths for $^{13}$CN Violet System (0,0) 
from Jenkins \& Wooldridge (1938).
The isotopic shift of the $^{13}$CN P$_1$($N''=1$) blend was obtained
from our observed spectrum.
Lines positions not listed in these papers were computed
using the molecular constants given by Prasad et al. and the standard
relations for the mass dependence of the various
molecular constants (cf Bernath 1995). 
The computed wavenumbers were
in good agreement with the laboratory measurements.
Predictions were extended to the isotopes $^{14}$CN and C$^{15}$N. 
Table~2 lists the wavelength shift of the various isotopes 
relative to CN. The isotopic shift is the same for
the three lines within one blend. 
For example: the $^{13}$CN $^{\rm R}$Q$_{21}$ (2) line
is at 3873.593 \AA.
In all cases the index of refraction of standard air was computed using
the formula given by Morton (1991).

The $f_{\rm V}$-value of the Violet System is well determined experimentally
and theoretically. Theoretical calculations by Knowles et al. (1988)
and Bauschlicher, Langhoff \& Taylor (1988) predict $f_{\rm V}$(0-0)=0.0345
and 0.0335 respectively. An earlier calculation by Larsson, Siegbahn
\& $\rm \AA$gren (1983) gave $f_{\rm V}$(0-0)=0.0324. The most accurate
experimental results are the measurements of the radiative
lifetime of the B$^2\Sigma^+$ state. Low rotational states of
the $v'=0$ level of the B
state have measured lifetimes (in ns) of $60.8\pm2.0$
(Luk \& Bersohn 1973), $65.1\pm0.8$ (Jackson 1974), and 
$66.7\pm1.4$ (Duric, Erman \& Larsson 1978).
Theoretical values (including the small contribution from
the B$^2\Sigma^+ \rightarrow $A$^2\Pi$) are 60.7 (Knowles et al.),
62.4 (Bauschlicher et al.) and 66.6 (Larsson et al.)
We adopt $f_{\rm V}(0-0)=0.033$ which is 
approximately the average of these various
theoretical and experimental estimates. The uncertainty is not
more than a few per cent.

\subsection{CN Red System}

Accurate wavelengths of Red System CN lines were taken from 
Davis \&
Phillips (1963), and the isotopic shifts for $^{13}$CN Red System
from Hosinsky et al. (1981).
Wavelengths not found in literature were computed
using the molecular constants of Prasad \& Bernath (1992) and
compared to those of the SCAN tape (J{\o}rgensen \& Larsson 1990).

Theory and experiment have not yet fully converged 
for the Red System
$f_{\rm R}$-value
- see brief review by Larsson (1994). 
Theoretical predictions are $f_{\rm R}(3-0)=3.34\times 10^{-4}$
(Knowles et al. 1988), $3.35 \times 10^{-4}$
(Bauschlicher et al. 1988), and $4.58\times 10^{-4}$
(Larsson et al. 1983).  Davis et al. (1986) measured the 
$f_{\rm R}$-values of six Red System bands relative to the Violet
(0-0) band using absorption lines produced by the transmission of
light through a column of CN in a furnace. Their result,
adjusted by 3\% to account for the difference between their adopted
and the above recommendation for $f_{\rm V}$(0-0), gives 
$f_{\rm R}(3-0)=(2.9\pm0.2)
\times 10^{-4}$, a value slightly smaller than the most recent 
theoretical calculations. Davis et al.'s results for
other bands were similarly smaller than the theoretical predictions.
Gredel, van Dishoeck \& Black (1991) analyze observations of CN Violet
(0-0) and Red (2-0) interstellar absorption 
lines in the same line of sight using the $f_{\rm R}$-values of
$f_{\rm V}(0-0)=0.0342$ and $f_{\rm R}(2-0)$ from Davis et al. (1986). Gredel et al.
remark that the Violet and Red lines give the same column density
but ``a slightly higher Red System oscillator strength would improve
the agreement between the Violet and the Red System data''.
Inspection of Gredel et al.'s results suggests that the theoretical
value given by Knowles et al. and Bauschlicher et al. fit the suggestion
of ``a slightly higher'' $f_{\rm R}$-value but Larsson et al.'s
value is probably too large.  There remains
an unresolved question posed by measurements of the radiative
lifetimes of vibrational levels of A$^2\Pi$ state. What
appear to be the best measurements,
radiative lifetimes from laser-induced fluorescence 
(Lu, Huang \& Halpern 1992),
are consistently shorter than the theoretical predictions:
the difference amounts to about 30 \% for $v'$=0 but a factor
3 for $v'$=7. These lifetimes imply that our adopted $f_{\rm R}$-values
may be underestimates.
We adopt the $f_{\rm R}$-value given by 
Knowles et al. and Bauschlicher et al.  with a small correction
for our adopted $f_{\rm V}$ value:
$f_{\rm R}(0-0)=23.7\times 10^{-4}$,
$f_{\rm R}(1-0)=19.1\times 10^{-4}$,
$f_{\rm R}(2-0)= 9.0\times 10^{-4}$,
$f_{\rm R}(3-0)= 3.3\times 10^{-4}$,  and
$f_{\rm R}(4-0)= 1.1\times 10^{-4}$.

\subsection{Oscillator strength of individual lines}

The oscillator strength of a transition is given to 
an acceptable precision by
$f_{N'J',N''J''}$ $= \left( \nu_{N'J',N''J''} / \nu_{0} \right) $
$f_{v'v''} S_{J'J''} / $ $\left( 2 J'' + 1 \right)$
where $\nu_{N'J',N''J''}$ and $\nu_{0}$ are the frequency of the transition
and the band origin respectively. $f_{v'v''}$ is the band
oscillator strength discussed in Sect 3.1 \& 3.2, and $S_{J'J''}$ is
the H\"{o}nl-London factor (also called the natural line strength).
The computation of the  H\"{o}nl-London factors is described in App.~A.

Almost all CN ``lines'' consist of two or three 
unresolved transitions. For example, the R-branch lines of the violet
system (Table~2 and Fig.~1) are effectively a triplet of
transitions ($\rm^{ }R_{1}$($N''$), $\rm^{ }R_{2 }$($N''$), 
and $\rm^{R}Q_{21}$($N''$)).
For the Red System, several branches provide an accompanying
satellite line (Table~3 and Fig.~2), 
e.g. , $\rm^{ }R_{2 }$($N''$) 
is accompanied by $\rm^{R}Q_{21}$ ($N''$).

The determination of column densities may use a curve of growth
computed for a single line. If unresolved components of a blend
are separated by more than about the Doppler width
$\Delta \lambda_{\rm D}$, the equivalent width of the lines
of individual transitions are added. If the separation
is much less than $\Delta \lambda_{\rm D}$, the blend may be 
presented as a single line with a so-called effective oscillator
strength 
(see App.~A). For separations comparable to $\Delta \lambda_{\rm D}$,
there is no simple way to represent the blend on the curve of growth
for a single line. 

\begin{figure} % Fig.~3
\centerline{\hbox{\psfig{figure=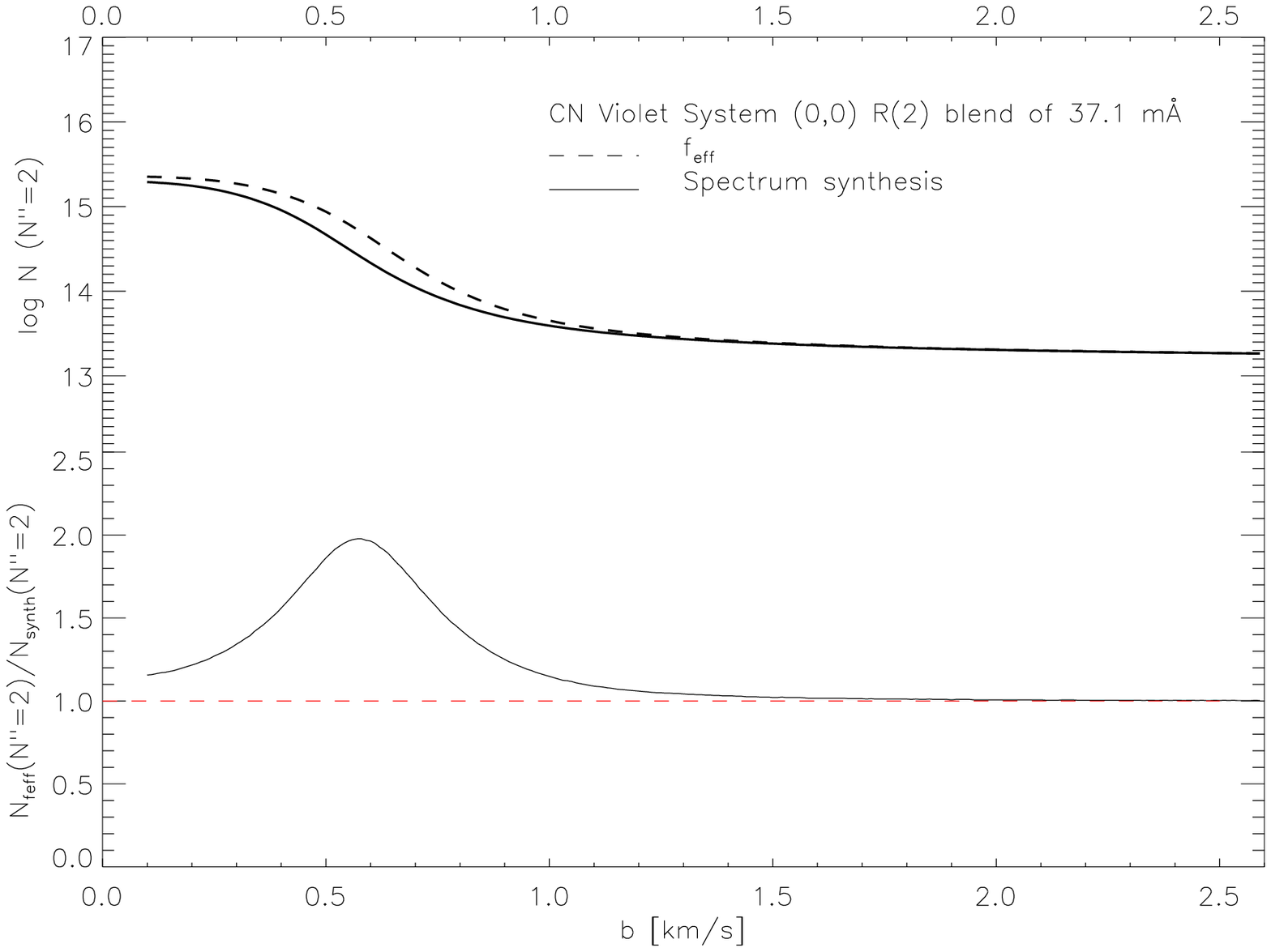,width=\columnwidth}}}
\caption{A comparison of the derived column density of the $N''=2$
level from the CN Violet System (0,0) R(2) line at 3873.37 \AA.
The solid line is obtained from the spectrum synthesis method, while
the dashed line uses the effective oscillator strength and a curve of
growth for a single, unblended line. The separation between the
two principal lines of the blend is 0.54 km~s$^{-1}$.} 
\end{figure}

A correct treatment of the overlap of spectral lines in a blend
necessarily requires the computation of synthetic spectra.
A computer program (MOLLEY-CN) was developed that computes
wavenumbers, H\"{o}nl-London factors and oscillator strength
for any (isotopic) band of the CN Red and Violet System. From these 
numbers a synthetic spectrum based on a Voigt profile is computed.
Input to the program is the rotational temperature, total column
density, isotope ratios, Doppler $b$-parameter,
and the spectral resolution of the output spectrum.
MOLLEY-CN is available on request from the authors.
As an aid to those who prefer the
method of effective oscillator strength, we have investigated
the error this latter method introduces on the derived column densities.
For this purpose we have chosen the CN Violet (0,0)
R(2) blend for which the two principal transitions are separated
by 0.54 km~s$^{-1}$. 
For a range of Doppler $b$-parameters the column density of the
$N''=2$ (F$_{1}$ and F$_{2}$) 
level has been computed using the effective oscillator
strength and synthetic spectra for a line with an equivalent width
of 37.1 m\AA. Fig.~3 demonstrates the results.
The derived $b$-parameter
in this work is $b=0.51\pm0.04$ km~s$^{-1}$, and the effective oscillator
strength method is clearly unacceptable for this work. We therefore
will use the method of computing synthetic spectra to determine the
column density responsible for the observed equivalent width of a 
``blend''.  The only remaining limitation is that the line might
be blended with transitions originating from other $N''$ level
and that the rotational temperature must be given to compute
such a blend. This limitation is easily overcome by iteration
using an improved estimate of the rotational temperature.

Fig.~3 can be qualitatively understood in the following way.
Use of $f_{\rm eff}$ assumes that the lines
which form a blend (R(2) is a triplet) have exactly the same
wavelength. This approximation
is valid only for $b$-values much larger than 
the separation between the lines of a blend. 
Fig.~3 demonstrates
this as the computed column density from the two methods
are the same for large $b$-values. 
For small $b$-values, the use of $f_{\rm eff}$ fails because
the lines are (partly) separated and the blend absorbs more
efficiently than assumed by the $f_{\rm eff}$ method. For
a given equivalent width of a line, this means that the
spectrum synthesis method yields a lower column density than
the $f_{\rm eff}$ method. 
The R(2) blend has three lines. For moderate column densities,
the contribution of the weakest line ($^{\rm R}$Q$_{21}$ (2)) 
can be neglected
and the ratio of the column density as derived from the two method
is given by the inverse of the slope of the curve of growth.
The slopes are: 1 for the linear part, 0.5 for the damping wings, and
$<$ 0.5 for the saturated part of the curve of growth, and
the maximum ratios are therefore: 1, 2 and $>$ 2. We see that
in Fig.~3 the ratio goes to 2 when the two strongest lines of
the triplet are on the saturated part of the curve of growth,
For larger column densities, the blend is highly saturated and
the broad damping wings contribute significantly to the equivalent
width. The damping wings will be the same in both cases
and the ratio will go to one. We note that in the ``real''
world lines never become this strong.

\section{Analysis}

\subsection{Velocities}

From the molecular lines with tabulated
laboratory wavenumbers, we find an average heliocentric 
radial velocity of $77.5\pm0.5$ km~s$^{-1}$.
To within the errors of the measurements, the velocity
is independent of the rotational level $N''$.
All velocities given in this article are corrected
to the heliocentric rest frame.
HD~56126 is a pulsating
star (Oudmaijer \& Bakker 1994, L\`{e}bre et al. 1996, 
discussion of Paper~II)
and the photospheric velocity derived from our spectra will
not be a good estimate of the systemic velocity of the star.
In order to derive an expansion velocity we use the 
systemic velocity derived from CO radio line emission (for 
overview see Paper I). For $v_{\ast,CO}=85.6\pm0.5$ km~s$^{-1}$~
we arrive at an expansion velocity of $v_{\rm exp}=8.1\pm0.7$ km~s$^{-1}$. 
This result is consistent with that derived from the width of
the CO lines: $v_{\rm exp}= 12.1 \pm 1.0$ km~s$^{-1}$ (Nyman et al. 1992). The 
expansion velocity is similar to expansion velocities
of AGB stars suggesting that the molecule-containing
layer around the post-AGB star,
which is possibly part of the terminal ``superwind'' of the AGB star,
was not ejected at very high velocity
(panel b of Fig.~3).

\subsection{Doppler $b$-parameter and synthetic spectra}

In paper~I it was demonstrated that the Red System (3,0) band
is just saturated while the stronger
(2,0) and (1,0) are saturated.
From this it is obvious that the
Violet System (0-0) band with $f_{\rm V}$-values a factor of ten higher 
than the Red System must be highly saturated. 
Of course, this circumstance facilitates the detection of the
$^{13}$CN lines. The degree of saturation can be judged
from Fig.~4 showing curves of growth for each $N''$ level.
Observed lines from a given $N''$ level are combined
into a single curve using the adopted $f$-values. The fitted
theoretical curve is computed from a single line assuming
$b=0.51$ km~s$^{-1}$. Clearly the unsaturated lines are all from the
Red System: 3-0 and 4-0 for the $N''=0$ level, but 
extending to the 1-0 band for the $N''=3$ level.
For all relevant $N''$ levels, the $^{12}$CN/$^{13}$CN ratio
is obtainable from a comparison of $^{13}$CN Violet System
lines with CN Red System lines of about the same strength.
Thus, the $^{12}$CN/$^{13}$CN ratio is insensitive to
the $b$-value but dependent on the ratio of the $f$-values
for the Violet and Red System. A sensitivity to the 
$b$-value does exist because, as noted above, the precise
curve of growth is dependent on the make-up of individual
CN features.
For a detailed analysis we must determine the Doppler $b$-parameter
and derive column densities from the observed equivalent widths
by taking into account the details of the blends and
optical depth effects.

For each transition (spectral line) there are two unknown parameters which
affect the observed equivalent width: the Doppler $b$-parameter
and the column density of the level the transition originates from ($N(N''$)).
The $b$-parameter must be the same for all lines, under the assumption
that they are all formed from the same gas. The $N(N'')$ is the same for
those transitions for a given level $N''$ of the same isotope.
In this study we included
$N''=0,1,2,3,4,5$ for CN and $N''=0,1,2,3$ for $^{13}$CN, thus
ten independent column densities.
We assumed that the F$_1$ and F$_2$ level population is
given by their degeneracy level and the energy of the average
$N''$ level.
We solved for $b$ and $N(N'')$ in the following manner: 
\newline
For a range of $b$-parameters ($0.10\leq b \leq 2.60$ km~s$^{-1}$~ in steps of 0.01 km~s$^{-1}$),
the column density responsible for the observed equivalent width of the blend
is determined by means of spectrum synthesis. We did not
fit the observed spectra directly, but instead used the observed equivalent
width. 
Although we have ten independent measure of the $b$-parameter, most
of the $N''$ levels are only probed by one or two lines and the
errors are rather large. We have limited the analysis therefore to
the CN $N''=0,1,2,3$ levels which are each probed by at least ten 
blends. For each $b$-parameter, $\log N(N'')$ and its standard deviation
where computed. The final $b$-parameter is given by the combination
for which the standard deviation is minimal, and 
by averaging the $b$-parameter
weighted by the number of lines. The
CN Doppler-parameter was found to be $b=0.51\pm0.04$ km~s$^{-1}$. This
same process was repeated for C$_{2}$ using the data of Paper I,
and this resulted in $b=0.53\pm0.03$ km~s$^{-1}$. Since the CN data is 
of higher quality we adopt $b=0.51\pm0.04$ km~s$^{-1}$~ for both CN and C$_{2}$.
Given the $b$ value, the $FWHM$ of the absorption line profile
should be at least $0.85\pm0.05$ km~s$^{-1}$. In order to resolve these lines 
a spectral resolution of $R\geq 350,000$ is needed. This confirms
that we are not be able to resolve these lines profiles in our spectra. 
The  line profiles of the synthetic
spectra can be perfectly fitted to the observed ones for this
$b$-value convolved to the spectral resolution of the observations.
No additional macroturbulent broadening is needed
to explain the observations.

\subsection{Column densities and  rotational temperatures}

Some additional lines were observed which are identified as
due to perturbations (Prasad et al. 1992). Perturbations
occur from the crossing of energy levels
in two different electronic manifolds (Kotlar et al. 1980).
The perturbations in the Violet System (0,0) band originate 
from B$^{2}\Sigma$ $v=0$ levels  perturbed by the A$^{2}\Pi$ $v=10$ level.
Perturbations affect the oscillator strength of lines.
Since a study of the perturbations is beyond the scope of
this work, we will not include the perturbed lines
in our analysis.

The derived average column density per $N''$ level and isotope is listed
in Table~4. 
Given these column densities an absolute rotational diagram
can be constructed (Fig.~5) and the rotational temperature
can be determined. Using a first order fit (constant $T_{\rm rot}$),  
we find $T_{\rm rot}($CN$)=12.8\pm0.6$ K,
and  $T_{\rm rot}(^{13}$CN$)=8.0\pm0.6$ K. 
The rotational temperature for $N''=0,1,2$ and 3 of CN
is determined very largely by the weak CN Red System lines,
especially the (3-0) and (4-0) lines
measured off our McDonald spectra: $T_{\rm rot}($CN$)=11.1\pm0.8$ K.
For $N''=4$ and 5, the column density is set almost
exclusively by a CN Violet System line which, in the case
of $N''=4$, is partially saturated: the $N''=4$ and 5 levels
give a temperature of ($T_{\rm rot}($CN$)=10.8\pm0.8$ K)
that within the errors is identical to the 
temperature from $N''=0$ to 3. The column density for $N''=4$ and
5 seems higher by 0.3 dex than given by extrapolation from 
$N''=0$ to 3. This offset
corresponding to a reduction of the Red System $f_{\rm R}$-values
by a factor two (relative to the Violet System) is probably
excluded by experimental and theoretical
results on the $f$-values (see above). A single rotational
temperature (Table~4) can be fitted to the points in Fig.~5.

Available $^{13}$CN lines, all from the Violet System, are
partially saturated except for the $N''=3$ lines. Saturation
means that the derived column densities are dependent on the
$b$-value and the adopted splitting of the unresolved lines that
make up the lines. In light of the fact that the $b$-values
from CN and C$_{2}$ are in excellent agreement, we believe
the $b$-value is not a serious contributor of uncertainty
to the determination of $T_{\rm rot}$ for $^{13}$CN. Certainly,
the column density of $N''=0$ cannot be lowered (relative to
$N''=3$) by 0.5 dex needed to make the $^{13}$CN rotational
temperature equal to that of CN.

From a study of millimeter emission lines (e.g. CO and CN) of seven
planetary nebulae, Bachiller et al. (1997b) found a kinetic temperature
of the CN line forming region of only 25~K, and a rotational
temperature of CN of 5-10~K. 
Quite unexpectedly we found different rotational temperatures
for CN and $^{13}$CN of $T_{\rm rot}=11.5\pm0.6$ and $8.0\pm0.6$ K
respectively. 

In light of the continuing discrepancies between experimental
and theoretical estimates for the CN Red System's $f_{\rm R}$-values
and the radiative lifetimes of the A$^{2}\Pi$ state, it is of interest
to consider if our spectra can give useful information on 
the ratio ($f_{\rm R}/f_{\rm V}$) of the Red to Violet System
$f$-values. Of course, the fundamental problem is that the 
Violet strongest lines of a given $N''$ are much stronger than the
strongest Red System lines of the same $N''$
($^{13}$CN lines are of comparable strength to the Red System CN lines
but the $^{12}$CN/$^{13}$CN ratio is not independently known).
A simple check of the $f_{\rm R}/f_{\rm V}$ is
possible by insisting that each band yields the same column
density. We have conducted
this exercise and found that relative to the
Violet System (0,0) band, the Red System bands give a larger
total column density by a factor 2.3 (1-0), 2.4 (2-0), 2.1 (3-0),
and 2.0 (4-0). To get consistent column densities the $f_{\rm R}$
values have to be decreased by this factor. This is clearly unacceptable
and we will adopt our initial estimated of the $f$-values to
proceed with our analysis.
It seems that the correction factor is wavelength dependent:
it increases with increasing wavelength. This could suggest
that there is scattered light (continuum, or CN  emission line radiation)
which fills in the absorption profile.
Since our observations were collected over several years
(the Red System was mainly observed in 1992 
and the Violet System only in 1997),
there is the possibility that the strength of the lines have
decreased with time. This idea can easily be tested by
simultaneously observing both systems.

\subsection{CN/$^{13}$CN/$^{14}$CN/C$^{15}$N isotope ratios}

Based on our findings of a different rotational temperature
for CN and $^{13}$CN, it is clear that the CN/$^{13}$CN
ratio will dependent on the $N''$
levels included in the determination of the column densities.
The CN/$^{13}$CN ratio for each N$''$  level is given in
Table~4 with an estimate of its uncertainty.
In order to determine
the isotope ratio of the line forming region we add
the contributions from all $N''$ levels:
CN/$^{13}$CN = $\sum_{N''=0}^{5} N(N'')^{\rm CN} 
/ \sum_{N''=0}^{3} N(N'')^{\rm ^{13}CN}=38.0\pm1.5$.
If the measured $T_{\rm rot}$ is applicable to $N'' \geq 5$ levels,
these bands contribute insignificantly to the summation.

Upper limits to the equivalent widths of 1.5 m\AA~ for both 
the $^{14}$CN and C$^{15}$N lines correspond to the following
lower limits: CN/C$^{15}$N($N''=1$) $\geq 2000$, 
and CN/$^{14}$CN($N''=1$) $\geq 2000$.
Assuming that the rotational temperature of $^{14}$CN and C$^{15}$N
is equal to the one of CN, we find lower limits
of  CN/C$^{15}$N $\geq 2000$, 
and CN/$^{14}$CN $\geq 2000$.

\begin{table*} % Table~4
\caption{Column densities and the isotope ratios derived from spectrum synthesis 
for $b=0.51$ km~s$^{-1}$, and the $f$-values given in Sect. 3.1 \& 3.2.}
\centerline{\begin{tabular}{llllllll}
\hline
\hline
$N''$ &\multicolumn{4}{c}{$\log N(N'')$ [cm$^{-2}]$}&CN/$^{13}$CN
                                                    &CN/$^{14}$CN
                                                    &CN/$^{15}$CN\\                                                   
\cline{2-5}
      &CN                &$^{13}$CN        &$^{14}$CN   &C$^{15}$N   &            &           &           \\
\hline
0     &$14.80\pm0.17$(10)&$13.22\pm0.10$(1)&            &            &$38.3\pm3.0$&           &           \\
1     &$15.05\pm0.17$(12)&$13.59\pm0.10$(2)&$\leq 11.72$&$\leq 11.72$&$28.8\pm2.1$&$\geq 2000$&$\geq 2000$\\
2     &$14.85\pm0.16$(17)&$13.16\pm0.10$(2)&            &            &$48.2\pm2.8$&           &           \\
3     &$14.39\pm0.13$(11)&$12.44\pm0.10$(2)&            &            &$89.2\pm3.9$&           &           \\
4     &$13.78\pm0.10$(01)&$\leq 11.79$     &            &            &            &           &           \\
5     &$12.70\pm0.10$(01)&                 &            &            &            &           &           \\
6     &$\leq 11.79$      &                 &            &            &            &           &           \\
      &                  &                 &            &            &            &           &           \\
      &                  &                 &            &            &\multicolumn{3}{c}{preferred ratios}\\
\cline{6-8}
Total:&$15.44\pm0.06$(52)&$13.86\pm0.09$(7)&$\leq 11.72$&$\leq 11.72$&$38.0\pm1.5$&$\geq 2000$&$\geq 2000$\\
$T_{\rm rot}$ [K]:&$11.5\pm0.6$&$8.0\pm0.6$&            &            &            &           &           \\  
\hline
\hline
\multicolumn{8}{l}{- numbers in brackets are the number of lines used
                     to determine the  average column density.} \\
\multicolumn{8}{l}{- errors are the standard deviations on the average, 
                     if less than 4 lines, 0.10 was adopted.} \\
\end{tabular}}
\end{table*}

\begin{figure*} % Fig.~4
\centerline{\hbox{\psfig{figure=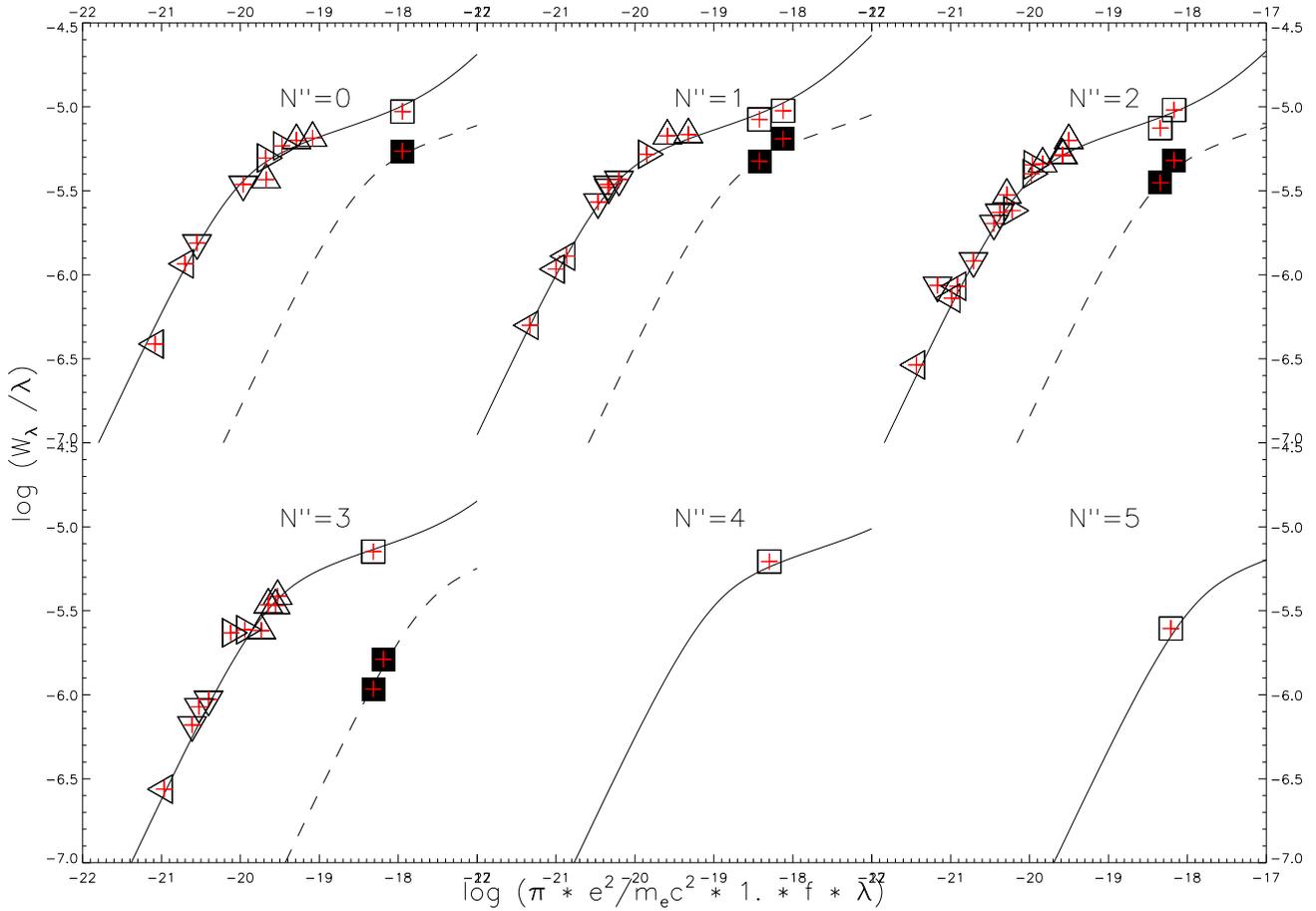,width=\textwidth}}}
\caption{Curve of growth for each $N''$ level. The displacement
of CN (solid line) and $^{13}$CN (dashed line) along the horizontal axis
gives the isotope ratio derived from that $N''$ level.
Squares for the Violet System (0,0),
and triangles for the Red System bands (triangle pointing
up (1,0), right (2,0), down (3,0), and left (4,0)).
The over plotted theoretical model is for $b=0.51$ km~s$^{-1}$.}
\end{figure*}

\begin{figure*} % Fig.~5
\centerline{\hbox{\psfig{figure=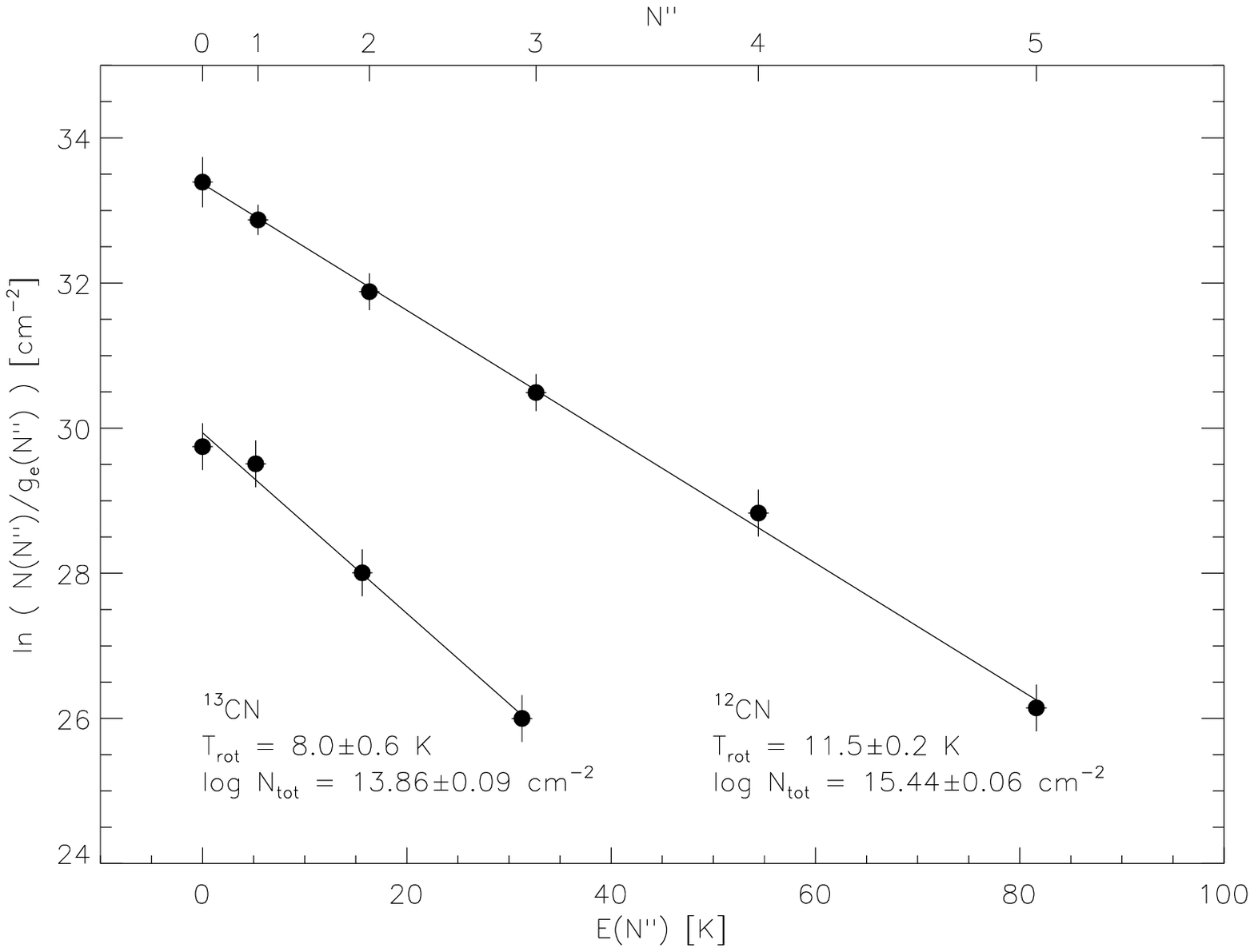,width=\textwidth}}}
\caption{Rotational diagram for CN and $^{13}$CN with
the column densities of Table~4 ($b=0.51$ km~s$^{-1}$). 
Note that the slope of the two
curves are different. This translates in CN having
a higher rotational temperature than $^{13}$CN.}
\end{figure*}

\section{Discussion}

Doppler parameter of  $b=0.51\pm0.04$ km~s$^{-1}$~
sets and upper limit on the kinetic temperature of 
$T_{\rm kin}=400\pm50$ K. This is rather high 
for gas which is about $10^{16}$ cm from the central star 
(Meixner et al. 1997).
Based on radio observations of CN towards
post-AGB stars, Bachiller et al. (1997b)
argue that they find $T_{\rm kin} \simeq 10 {\rm ~to~} 25$ K.
The $b$-parameter must, therefore, have an additional contribution from
turbulence. The CN rotational
temperature cannot be used to estimated the kinetic temperature
since the molecule is sub-thermal. C$_2$ behaves supra-thermal,
but the rotational temperature derive from the $J''=0$ and $J''=1$ level
is a good indicator for the gas kinetic temperature. This
would suggest $T_{\rm kin} \simeq 189$~K (Bakker \& Lambert 1997). This value
is also much higher than would be expected, and it is clear that 
we need an independent accurate determination of the kinetic 
temperature. The  turbulence required 
is in the range of $v_{\rm micro}=0.36 {\rm ~to~} 
0.50$ km~s$^{-1}$~ and depends on the adopted $T_{\rm kin}$.
The source of turbulence is not identified, but might
be the result of small scale turbulence,
non-Maxwellian velocity distribution,
velocity stratification, the presence of multiple unresolved
absorption components due to separate clumps.

% selective isotope depletion
An isotopic ratio as determined from 
molecules ($^{i} {\rm XY}$ and $^{j} {\rm XY}$)
does not necessarily reflect the intrinsic
isotopic ratio ($^{i} {\rm X}/^{j} {X}$)
of the gas. 
For example, interstellar (and probably circumstellar) CO
is subject to two competing processes 
which affect this ratio: photodissociation by
line radiation and ion-molecule charge exchange
(Watson et al. 1976). 
The net result depends on the local conditions (density, temperature,
and radiation field).

Since photodissociation
for the CO molecule is driven by line radiation,
the more abundant CO,
 the more the molecule can shield itself against
the dissociating by photons. The CN molecule is 
photodissociated by continuum radiation, 
not by lines. 
The ion-molecule reaction on the other hand will operate
for CN.
The gas-phase ion-molecule reaction for CN is given by:

\begin{eqnarray}
{\rm ^{13}C^+} + {\rm ^{12}CN}   \rightleftharpoons 
{\rm ^{12}C^+} + {\rm ^{13}CN} + \Delta E 
\end{eqnarray}

\noindent
where $\Delta E$ is the zero-point
energy difference derived from the molecular parameters (Prasad et
al. 1992) and the relation of isotopic dependence of the Dunham 
coefficients. We find 31, 58, and 23~K for $^{13}$CN, $^{14}$CN,
and C$^{15}$N respectively.
The reaction to the right will be referred to as the forward
reaction ($k_{f}$) and to the left as the reverse reaction ($k_{r}$). 
C$^{+}$ can be produced by cosmic ray ionization
of free carbon atoms and molecules containing carbon
(e.g. CO), and by the reactions of He$^{+}$
with these molecules  (where CO is the dominant one).

Our picture of the chemistry of the CN molecule in the
circumstellar shell is a simple one. CN molecules which
are produced by photodissociation of the parent molecule HCN, 
are subsequently destroyed by photodissociation. Between their
formation and destruction, CN molecules participate in the 
isotopic exchange reaction.
If the time scale for the isotopic exchange reaction is shorter
than the lifetime of CN, then the exchange reaction will
reach equilibrium:

\begin{eqnarray}
R_{\rm eq}=  R_{0} / \alpha \\
{\rm with} \nonumber  \\
\alpha    =\frac{k_{f}}{k_{r}} = e^{ \Delta E/k T_{\rm kin}}  \\
R_{0     }=\frac{[^{12} {\rm C} ^{+}]}{[^{13} {\rm C}^{+}]} =
           \frac{[^{12} {\rm C}     ]}{[^{13} {\rm C}    ]} \\ 
R_{\rm eq}=\frac{[^{12} \rm CN]}{[^{13} \rm CN]}
\end{eqnarray}

\noindent
with the square bracket denoting the local density of that species
in cm$^{-3}$, $R_{\rm eq}$ and $R_{0}$ the equilibrium CN and
true C isotope ratio.
A lower limit to the kinetic temperature
is given by the CN rotational temperature of 12~K. This
gives an upper limit on $\alpha$ of 13. For higher temperature,
$\alpha$ decreases to unity. This clearly indicates that
if the ion-molecule reaction is in equilibrium, we
may underestimate the isotope ratio by 
as much as a factor 13! For a more realistic $T_{\rm kin}
\simeq 25$ K, and an $R_{\rm eq}=23$, we find a true
$R_{0}={\rm C}/^{13}{\rm C}= 80$.

To assess the behavior of the ratio $\rm ^{12} CN/^{13}CN $ in
an expanding shell, we use the theoretical work
of Cherchneff et al. (1993) on the chemistry of
the circumstellar shell surrounding the carbon star
IRC~+10216. The shell radius where
CN is more abundant than half its 
peak abundance are between $0.7\times 10^{17}$ and $1.3\times 10^{17}$ cm. 
Together with the  expansion velocity this gives a lifetime for CN of 
about $2\times 10^{3}$ years. Basically this is the lifetime of a single
CN molecule after photodissociation of HCN, before 
the photodestruction of CN, not taking into account any
other reaction that could produce or destroy CN.

The time scale for the ion-molecule reaction is more difficult
to estimate. Adams et al. (1985) computed reaction rates ($k_{f}$) for 
polar molecules and found that they are as high as $10^{-7}$ 
cm$^{3}$ s$^{-1}$ for low temperature gas ($T_{\rm kin}<50 $ K). 
The radius at which the CN abundance peaks ($r \simeq 1.0\times 10^{17}$
cm$^{-3}$)
has a C$^{+}$ abundance of $3\times10^{-4}$ cm$^{-3}$. The time scale
for the isotopic exchange reaction as derived from the
forward rate coefficient and this C$^{+}$ particle density
is of the order of $1\times 10^{3}$ years. The isotopic exchange reaction time scale
is comparable to the CN life time. This suggest that we have to look
into this process in some greater detail.

\begin{figure*} % Fig.~6
\centerline{\hbox{\psfig{figure=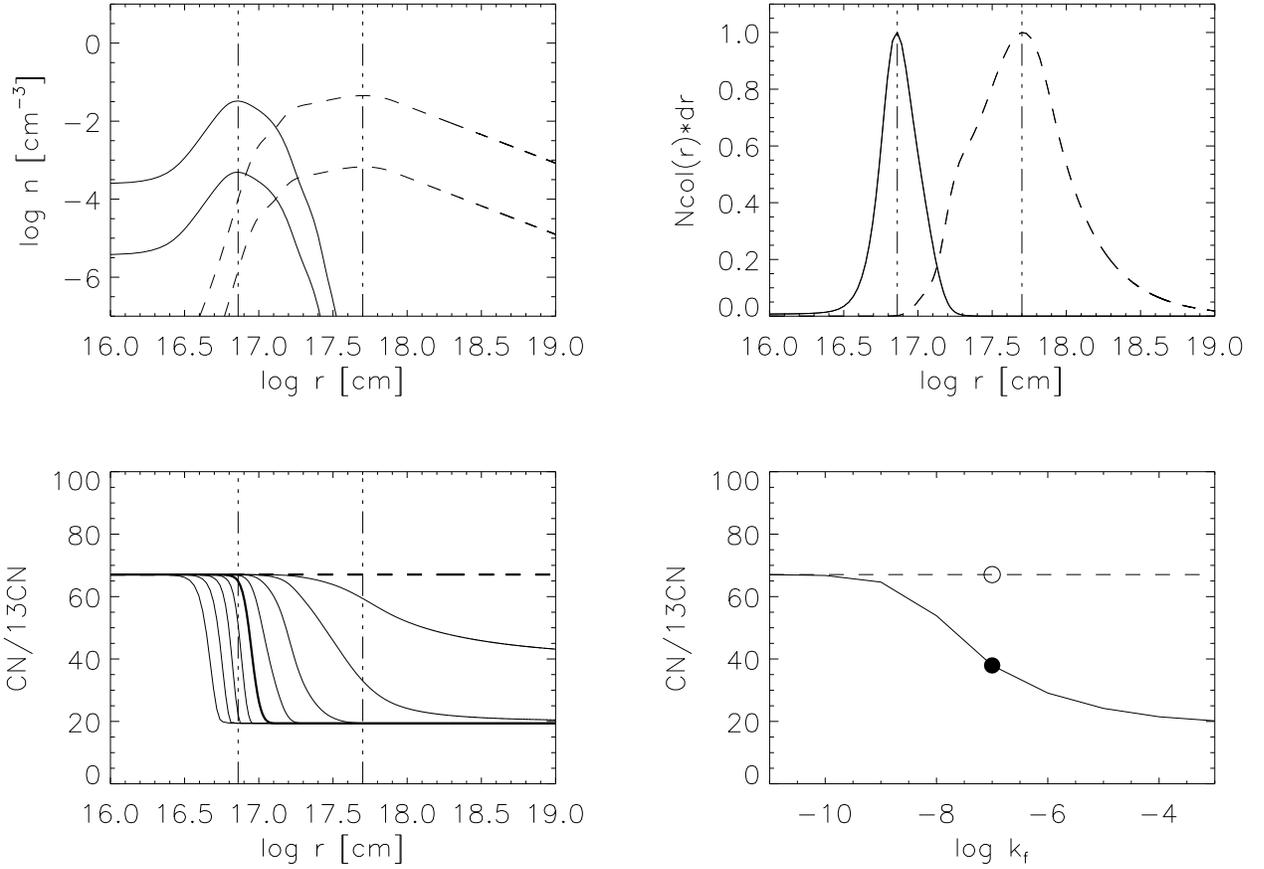,width=\textwidth}}}
\caption{(a) The density of CN and $^{13}$CN (solid lines) and
C$^{+}$ and $^{13}$C$^{+}$ (dashed lines) as function of distance
to the star (adopted from Cherchneff et al. 1993). (b) The contribution
of the column density to the total column density, normalized
to a peak value of 1. (c) The local isotope ratio for CN and C$^{+}$
as function of radius for reaction rates of $\log k_{f}$=-11 to -3
in steps of 1. $k_{f}=1.0 \times 10^{-7}$ is enhanced. (d) The
final observed isotope ratio, with a dot at our preferred 
rate coefficient.}
\end{figure*}

The effect of the isotopic exchange reaction should ideally be
studied by computing self consistent chemical models for the 
abundances and radiation field appropriate for HD~56126.
Such a study is beyond the scope of this work, instead we 
will make some estimates. A model of the circumstellar shell of
the carbon star IRC~+10216 has been computed by
Cherchneff et al. (1993). This model uses a 
large number of chemical  reactions, the interstellar
radiation field, circumstellar reddening, and computes the 
abundances of the most important molecular species.
The two major differences between IRC~+10216 and HD~56126 are
(i) the inner shell radius of HD~56126 will be much larger than
for IRC~+10216, and (ii) the radiation field of HD~56126 (7000 K,
Klochkova 1995) is significantly hotter than that of IRC~+10216
(3500 K). The larger inner radius
does not affect the chemical structure of the remaining shell.
A 7000~K stellar radiation field  has fewer UV photons than the interstellar
UV field (see Paper~II) such that the interstellar radiation
field will dominate the photodissociation processes. Therefore
the chemical structure of IRC~+10216 is applicable to HD~56126.
From the model, we infer that CN is produced mainly
by photodissociation of HCN and destroyed by photodissociation
into C and N.
From Cherchneff et al. we take the CN and C$^{+}$ density
(panel a of Fig.~6). Assuming that the isotopic exchange reaction
for CN has reached equilibrium ($T_{\rm kin} =25$ K), we
assume a true isotope ratio of $R_{0} = 67$.
From this we compute where the contribution
to the observed column density comes from (panel b of Fig.~6):
the line is formed where the CN density is highest.
For a range of reaction rates we have computed the local
CN and $^{13}$CN densities, while the C$^{+}$ density in unaffected
by the isotopic exchange reaction.
(panel c of Fig.~6).  C$^{+}$ is mainly formed from CO and
the processes involved in the production of C$^{+}$ are not
coupled to CN. The assumed reaction rate is $k_{f}=1.0 \times
10^{-7}$. We see that the transition from
the initial CN isotope ratio to the equilibrium value occurs
where the CN density is highest and that this transition is rather
fast. From these densities, we can computed the predicted observed
CN isotope ratio (panel d of Fig.~6). To obtain an observed
CN isotope ratio of 38 at $T_{\rm kin}=25$ K, 
we have to start with an intrinsic
isotope ratio of $R_{0}=67$ and  $R_{\rm eq} = 19$. 
Clearly, the isotopic
exchange reaction is important.
Based on this analysis we suggest that the intrinsic carbon isotope
ratio is $^{12}$C/$^{13}$C=$67$ for $T_{\rm kin} = 25$~K.
Adams et al. (1985) show that the rate coefficient is dependent on the
rotational quantum number. The higher the rotational
quantum number, the lower the rate coefficient. 
If this dependence is rather strong, then the isotopic exchange reaction
would be able to explain the difference in rotational temperature between
CN and $^{13}$CN: for $N''=0$ more $^{13}$CN is formed than for $N''=5$
leading to a lower CN isotope ratio as derived from $N''=0$ than
from $N''=4$. 

Isotopic exchange reactions are very efficient at low temperatures 
and for high dipole moment molecules.
From this we predict that CO (with a lower dipole moment) will be less
affected, while C$_{2}$ with no dipole moment is unaffected. 
C$_{2}$ will therefore give the true isotope ratio and combined with
CN this may give the kinetic temperature of the gas.
The isotopic exchange reaction also enhances the abundance
of $^{14}$CN and C$^{15}$N.

% mass-loss rate
In order to derive the mass-loss rate from the CN absorption lines 
we adopt the CN abundance (in number of particles)
of $X_{\rm CN}=1.0 \times 10^{-6}$ for
AFGL~2688 (Bachiller et al. 1997a). AFGL~2688 (the Egg Nebula)
is so far the only C2CN stars for which radio emission of CN has been observed.
The inner and outer radius of the ejecta of HD~56126 have been determined
by Meixner et al. (1997) by modeling the spatial distribution of 
the infrared source: $r_{\rm inner}= 45\pm2 \times 10^{15}$ cm
(1.0$\pm$0.15''),
and $r_{\rm outer}=155\pm15 \times 10^{15}$ cm 
($3.6\times3.3''$ at 11.8 $\mu$m).
Assuming that CN coincides with the infrared source, we find
$\dot M=4.5\pm0.8\times 10^{-5}$ M$_{\odot}$ yr$^{-1}$. 

%implications of new data

The prototype of the AGB stars, IRC~+10216/CW Leo,
is a massive, highly-evolved carbon star 
with $3 M_{\odot} \leq M_{\rm ZAMS} \leq 5 M_{\odot}$
(Gu\'{e}lin et al. 1995). For IRC~+10216 the isotope
ratios are well determined and give an estimate of the ratios
which one might expect to detect for a carbon-rich post-AGB star
like HD~56126: 
C/$^{13}$C=44$^{+3}_{-3}$, 
$^{12}$C/$^{14}$C$\geq 62600$,
N/$^{15}$N$\geq$5300, 
O/$^{17}$O=840$^{+230}_{-170}$,
and 
O/$^{18}$O=1260$^{+315}_{-240}$ (see Forestini \& Charbonnel for 
an overview). 
Our estimate of the $^{12}$C/$^{13}$C ratio for HD~56126's shell is 
consistent with that of IRC~+10216. Results for circumstellar
shells around four other carbon stars were provided by Kahane et al. (1992):
C/$^{13} \leq 60$, $ >30$, $32^{+10}_{-7}$ and $31^{+6}_{-5}$. These too suggest
that HD~56126 is not exceptional.  

A comparison may also be made with isotopic ratios pertaining to the
photosphere of cool carbon (AGB) stars. It must be borne in mind
that the photospheric ratios 
may evolve before the stars shed the
material that becomes the circumstellar shell of a post-AGB star. Luminous
AGB stars of intermediate mass experience H-burning at the base of their
convective envelope. This decreases the C/$^{13}$C ratio but rather
quickly also converts the star from C-rich to O-rich. It seems
improbable that C-rich post-AGB stars will have evolved from this
intermediate stage. At lower luminosity and for lower mass stars, 
$^{12}$C from the He-burning shell is added following a thermal pulse.
Repetition of this addition will increase the photosphere's C/$^{13}$C ratio
unless H-burning converts $^{12}$C to $^{13}$C. On this scenario, the
C/$^{13}$C ratio of a post-AGB star might be higher than that of a typical
AGB progenitor.   

Through analysis of infrared spectra of N-type carbon stars,
Lambert et al. (1986) showed that most stars had a C/$^{13}$C ratio
in the range 30 to 70: a value of 67 is not unusual. Ohnaka \& Tsuji (1996),
however, from analysis of spectra near 8000 \AA reported, systematically
lower ratios but a ratio of 67 would not be unusual, although on the high
end of the distribution function. This possible shift between the C/$^{13}$C
ratio of stars and HD~56126 may be due to the fact that the stellar
ratios continue to evolve before the post-AGB phase is reached. On the other
hand, quantitative analysis
of SiC circumstellar grains extracted from meteorites shows a C/$^{13}$C
distribution function that mimics well Lambert et al.'s stellar
distribution, which suggests, perhaps, that evolutionary changes are minimal.

A priority goal has to be the determination of more accurate C/$^{13}$C ratios
for HD~56126 and an adequate sample of post-AGB stars. In the case of
HD~56126 the ratio is presently uncertain for several reasons. A
minor contributor to uncertainty is the combined use of CN Violet
and Red System lines and the presence of 
uncertainty over the $f$-values of the 
Red System. This uncertainty could be avoided by re-observation of the
strongest CN Red System bands to higher S/N ratio such that their $^{13}$CN
lines are detectable. A key problem will remain in that the CN molecules
are fractionated in the shell and this has to be modeled in order to
extract the C/$^{13}$C ratio that is of primary interest. This
extraction requires a better understanding of physical conditions in the
shell. We suggest that this understanding may be obtained by analyzing
spectra of other C-containing molecules. In particular, C$_2$ is
accessible by the Phillips system, and CO is probably detectable in
the infrared. Lacking an electric dipole moment, the isotopic exchange
reaction involving C$_2$ will be very slow, and, as we have already
shown observationally (Bakker \& Lambert 1997), the C$_2$ and CN rotational
ladders are influenced differently by the local density and the ambient
radiation field. The CO molecule has a smaller dipole moment than CN, and
its photodissociation is line-dominated. Measurements of the ratios
C$_2$/$^{12}$C$^{13}$C, and CO/$^{13}$CO together with the CN/$^{13}$CN
ratio will shed light on the physical conditions in the shell, and on the
true $^{12}$C/$^{13}$C ratio. If such data can be assembled for a sample
of post-AGB stars, it will be possible to trace a link with their AGB 
antecedents.

\acknowledgments
The authors acknowledge the support of
the National Science Foundation (Grant No.
AST-9618414) and the Robert A. Welch Foundation of Houston, Texas.
We thank Joe Smreker for assistance at the telescope,
and Eric Herbst for a discussion on the isotopic exchange reaction.
This research has made use of the Simbad database, operated at
CDS, Strasbourg, France, and the ADS service.

\appendix
\onecolumn

\section{H\"{o}nl-London factors and effective oscillator strength}

H\"{o}nl-London factors were computed from the equations given in
Schadee (1964)  and 
Kov\'{a}cs (1969), and normalized according to the sum rule
$\sum_{J'} S_{J'J''} = \left( 2 - \delta_{0,\Lambda'+\Lambda''} \right) \times
\left( 2 S'' + 1 \right) \times \left( 2 J'' + 1 \right)$ as stated by
(Larsson 1983).
For both the Red and Violet System we take $S''=0.5$ since
both originate from the same ground level ($^2\Sigma$).
For the Red System ($\Pi-\Sigma$, $\Lambda'=1$ and $\Lambda''=0$)
we find $\delta_{0,1}=0$ and for the Violet System 
($\Sigma-\Sigma$, $\Lambda'=0$ and $\Lambda''=0$)
$\delta_{0,0}=0$.
For the Red System we summed over twelve transitions
following the selection rule ($\Delta J =-1,0,1$ and $\Delta N =-2,-1,0,1,2$)
(six for $N''=J''+0.5$ and six for $N''=J''-0.5$ for a given $J''$)
($\rm^{ }P_{1 }$,
 $\rm^{ }Q_{1 }$,
 $\rm^{ }R_{1 }$,
 $\rm^{Q}P_{21}$,
 $\rm^{R}Q_{21}$,
 $\rm^{P}Q_{12}$,
 $\rm^{Q}R_{12}$,
 $\rm^{ }P_{2 }$,
 $\rm^{ }Q_{2 }$,
 $\rm^{ }R_{2 }$,
 $\rm^{S}R_{21}$ and 
 $\rm^{O}P_{12}$)
and for the Violet System over six transitions
following the selection rule ($\Delta J =-1,0,1$ and $\Delta N =-1,1$  )
(three for $N''=J''+0.5$ and three for $N''=J''-0.5$ for a given $J''$)
($\rm^{ }P_{1 }$,
 $\rm^{ }R_{1 }$,
 $\rm^{Q}P_{21}$,
 $\rm^{Q}R_{12}$,
 $\rm^{ }P_{2 }$ and
 $\rm^{ }R_{2 }$).
For both the Red and Violet System this results in 
the relation:  $\sum_{J'} f_{J'J''} = 2 \times f_{v'v''}$.
The oscillator strength of individual lines were checked against
those of the SCAN tape (for the Red system) and those listed
in literature (for the Violet system).

Given a Doppler broadening of $b=0.51$ km~s$^{-1}$, 
transitions which are separated from each other
by less than 0.012 \AA~ (Violet System), and 0.022 \AA~ (Red System),
overlap and the two transitions should be treated as one line
with an effective oscillator strength. Lines which are
separated by more than this interval should be treated separately,
although they might not be resolved in the spectrum.
In physical terminology: if lines are separated by less than the 
Doppler $b$-parameters, their optical depth
at any frequency point  should be added before
computing the line profile ($I=e^{-(\tau_{1}+\tau_{2})}$). 
If lines are separated by more than
the Doppler $b$-parameter, the line profiles at 
any frequency point can be added
($I=0.5 \times  e^{-\tau_{1}}+ 0.5 \times e^{-\tau_{2}}$), 
although adding the optical depth is also fine.
For each blend we can determine an effective oscillator strength
in the following way:

The F1 ($J''=N''+0.5$) and F2 ($J''=N''-0.5$)
levels have different degeneracy: 

\begin{eqnarray}
g_e(F1)=2 \times J''+1=2 \times N''+2 \\
g_e(F2)=2 \times J''+1=2 \times N'' 
\end{eqnarray}

\noindent
Each $N''$ levels has therefore a degeneracy of 
\begin{eqnarray}
g_e(F1+F2)=2 \times (2 \times N''+1)
\end{eqnarray}

This is even valid for $N''=0$ although that level has no F2 levels.
The reason for
this is that each  $J''$ level is split into $2 \times J''+1$ so called $m$-levels. 
Those levels are only separated from each other under the influence of
a strong external magnetic field (Zeeman splitting). In the case of
circumstellar CN there is no observational evidence to assume that these
levels are separated which indicates that there is no measurable magnetic
field in the line forming region.
To first order
the $m$ levels for F1 and F2 for a given $N''$ have the same energy and therefore
the same population. This allows us to write: \newline

\begin{eqnarray}
N(N'',F1)=g_e(F1)/g_e(F1+F2) \times N(N'',F1+F2)  \\
N(N'',F2)=g_e(F2)/g_e(F1+F2) \times N(N'',F1+F2)  \\
N(N'',F1+F2)=N(N'',F1)+N(N'',F2)           \\
{\rm and } \nonumber \\
f_{\rm eff} N(N'') = \sum_{i} f_{i} N_{i}(N'')\\
f_{\rm eff} N(N'') = \left( \sum_{i} f_{i}(F1) \times (N''+1) 
                          + \sum_{i} f_{i}(F2) \times N'' \right)
 \times
\frac{N(N'')}{(2 \times N''+1) }
\end{eqnarray}

\noindent
As examples we take two blends.
$\Delta N=N' - N''= 0$ at 6194.5 \AA~ (Red System):
\begin{eqnarray}
f_{\rm eff}N(N''=1)=\left( 0.091 \times 2 + 0.363 \times 1 \right) 
   \times
   \frac{N(N''=1)}{ 3 } = 0.182 N(N''=1) \nonumber
\end{eqnarray}

\noindent
and $\Delta N =N' - N''=-1$ at 3876.3 \AA~ (Violet System):
\begin{eqnarray}
f_{\rm eff}N(N''=2)=\left(0.0132 \times 3 + 0.0022 \times 2 +
   0.0110 \times 2  \right)  \times
  \frac{ N(N''=2) }{ 5} = 0.0132  N(N''=2) \nonumber
\end{eqnarray}

\section{Extra line lists}

\begin{table*}  % Table~8
\caption{Index of refraction for air.
Table will not appear in journal.}
\begin{tabular}{ll}
\hline
\hline
Band                &n$_{\rm air}$ \\
\hline
Violet System (0,0) & 1.000283     \\
Red System (1,0)    & 1.000274     \\
Red System (2,0)    & 1.000275     \\
Red System (3,0)    & 1.000276     \\
Red System (4,0)    & 1.000277     \\ 
\hline
\hline
\end{tabular}
\end{table*}

\begin{table*} % Table~5, not in article
\caption{CN Red System (1,0) with $f_{(1,0)}=1.90\times 10^{-4}$.
Derived column densities are given in Table~4.
Table will not appear in journal.}
\centerline{\begin{tabular}{rllllll}
\hline
\hline
$B(N'')$        &$\lambda_{\rm rest}$ [\AA]&$f_{N'J',N''J''}\times10^{4}$&
$f_{\rm eff}\times 10^{4}N(N'')$ [cm$^{-2}$]&$W_{\lambda}$ [m\AA]&
Remark \\ 
\hline
$\rm ^{S}R_{21}$(3)&9129.142c& 1.75& 1.00N(3)&no    &  \\
\cline{1-4}
$\rm ^{S}R_{21}$(2)&9133.730c& 1.95& 1.17N(2)&no    &  \\
\cline{1-4}
$\rm ^{S}R_{21}$(1)&9136.579c& 2.19& 1.46N(1)&no    &  \\
\cline{1-4}
$\rm ^{S}R_{21}$(0)&9139.686c& 2.62& 2.62N(0)& 33.7 &  \\ 
\cline{1-4}
$\rm ^{ }R_{2 }$(3)&9141.169c& 3.55&         &      &  \\
$\rm ^{R}Q_{21}$(3)&9141.189c& 3.77& 3.67N(3)& 35.3 &  \\
 \cline{1-4}
$\rm ^{ }R_{2 }$(2)&9141.870c& 3.55&         &      &  \\
$\rm ^{R}Q_{21}$(2)&9141.886c& 4.11& 3.89N(2)& 57.8 &  \\
 \cline{1-4}
$\rm ^{ }R_{2 }$(1)&9142.828c& 3.76&         &      &  \\
$\rm ^{R}Q_{21}$(1)&9142.838c& 4.62& 4.33N(1)& 57.8 &polluted\\
 \cline{1-4}
$\rm ^{R}Q_{21}$(0)&9144.042c& 6.34& 6.34N(0)& 57.8 &  \\
\cline{1-4}
$\rm ^{ }Q_{2 }$(1)&9147.208c& 6.34& 3.18N(1)& 61.5 &  \\
$\rm ^{Q}P_{21}$(1)&9147.217c& 1.60&         &      &  \\
\cline{1-4}
$\rm ^{ }Q_{2 }$(2)&9149.167c& 5.53& 3.29N(2)& 47.0 &  \\
$\rm ^{Q}P_{21}$(2)&9149.182c& 1.79&         &      &  \\
\cline{1-4}
$\rm ^{ }Q_{2 }$(3)&9151.381c& 5.67& 3.45N(3)& 31.1 &  \\
$\rm ^{Q}P_{21}$(3)&9151.401c& 1.79&         &      &  \\
\cline{1-4}
$\rm ^{ }P_{2 }$(2)&9153.532c& 1.59&0.636N(2)& 27.4 &  \\
\cline{1-4}
$\rm ^{ }P_{2 }$(3)&9158.658c& 2.02&0.866N(3)&      &  \\
\cline{1-4}
$\rm ^{ }R_{1 }$(3)&9177.032c& 4.86& 2.78N(3)& 31.5 &  \\
\cline{1-4}
$\rm ^{ }R_{1 }$(2)&9179.911c& 5.32& 3.19N(2)& 47.6 &  \\
\cline{1-4}
$\rm ^{ }R_{1 }$(1)&9183.209c& 6.38& 4.25N(1)&101.8 &polluted\\
\cline{1-4}
$\rm ^{ }R_{1 }$(0)&9186.932c&10.10&10.10N(0)& 59.9 &  \\
\cline{1-4}
$\rm ^{Q}R_{12}$(3)&9189.465c& 3.72&         &      &  \\
$\rm ^{ }Q_{1 }$(3)&9189.486c& 5.61&         & 97.8 &polluted\\
$\rm ^{Q}R_{12}$(2)&9189.581c& 5.03&         &      &  \\
$\rm ^{ }Q_{2 }$(2)&9189.597c& 5.14&         &      &  \\
\cline{1-4}
$\rm ^{Q}R_{12}$(1)&9190.120c& 8.96& 5.83N(1)& 62.9 &  \\
$\rm ^{ }Q_{1 }$(1)&9190.129c& 4.26&         &      &  \\
\cline{1-4}
$\rm ^{P}Q_{12}$(2)&9196.511c& 3.38&1.802N(2)& 42.2 &  \\
$\rm ^{ }P_{1 }$(2)&9196.525c& 0.75&         &      &  \\
\cline{1-4}
$\rm ^{P}Q_{12}$(3)&9199.171c& 3.59& 2.26N(3)& 22.2 &  \\
$\rm ^{ }P_{1 }$(3)&9199.192c& 1.27&         &      &  \\
\cline{1-4}
$\rm ^{O}P_{12}$(3)&9206.114c& 0.52&0.223N(3)&no    &  \\
\hline
\hline
\multicolumn{5}{l}{no: not observed thus not available} \\
\end{tabular}}
\end{table*}

\begin{table*} % Table~6 , not in article
\caption{CN Red System (2,0) with $f_{(2,0)}=9.03\times 10^{-4}$.
Derived column densities are given in Table~4.
Table will not appear in journal.}
\centerline{\begin{tabular}{rllllll}
\hline
\hline
$B(N'')$        &$\lambda_{\rm rest}$ [\AA]&$f_{N'J',N''J''}\times10^{4}$&
$f_{\rm eff}\times 10^{4}N(N'')$ [cm$^{-2}$] &$W_{\lambda}$ [m\AA]&
Remark \\ 
\hline
$\rm ^{S}R_{21}$(0)&7871.654c&1.24&1.24N(0)& 29.4 &polluted\\
\cline{1-4}
$\rm ^{ }R_{2 }$(3)&7872.889c&1.68&        &      &\\
$\rm ^{R}Q_{21}$(3)&7872.905c&1.79&1.74N(3)& 18.5 &polluted\\
\cline{1-4}
$\rm ^{ }R_{2 }$(2)&7873.332c&1.68&        &      &\\
$\rm ^{R}Q_{21}$(2)&7873.343c&1.96&1.85N(2)& 38.8 &polluted\\
\cline{1-4}
$\rm ^{ }R_{2 }$(1)&7873.985c&1.79&        &      &\\
$\rm ^{R}Q_{21}$(1)&7873.992c&2.19&2.06N(1)& 45.8 &polluted\\
\cline{1-4}
$\rm ^{R}Q_{21}$(0)&7874.852c&3.01&3.01N(0)& 39.1 &\\
\cline{1-4}
$\rm ^{ }Q_{2 }$(1)&7877.198c&3.01&1.51N(1)& 98.2 &polluted\\
$\rm ^{Q}P_{21}$(1)&7877.205c&0.76&        &      &\\
\cline{1-4}
$\rm ^{ }Q_{2 }$(2)&7878.686c&2.62&1.56N(2)& 35.6 &\\
$\rm ^{Q}P_{21}$(2)&7878.697c&0.85&        &      &\\
\cline{1-4}
$\rm ^{ }Q_{2 }$(3)&7880.384c&2.69&1.43N(3)& 19.2 &\\
$\rm ^{Q}P_{21}$(3)&7880.400c&0.85&        &      &\\
\cline{1-4}
$\rm ^{ }P_{2 }$(2)&7881.889c&0.76&0.30N(2)& nd   &\\
\cline{1-4}
$\rm ^{ }P_{2 }$(3)&7885.725c&0.96&0.41N(3)& nd   &\\
\cline{1-4}
$\rm ^{ }R_{1 }$(3)&7899.481c&2.31&1.35N(3)& 42.0 &polluted\\
\cline{1-4}
$\rm ^{ }R_{1 }$(2)&7901.520c&2.52&1.51N(2)& 31.6 &\\
\cline{1-4}
$\rm ^{ }R_{1 }$(1)&7903.892c&3.03&2.02N(1)& 41.3 &\\
\cline{1-4}
$\rm ^{ }R_{1 }$(0)&7906.598c&4.79&4.79N(0)& 46.3 &\\
\cline{1-4}
$\rm ^{Q}R_{12}$(3)&7908.597c&1.76&        &      &\\
$\rm ^{Q}R_{12}$(2)&7908.611c&2.39&        & 50.1 &polluted\\
$\rm ^{ }Q_{1 }$(3)&7908.613c&2.66&        &      &\\
$\rm ^{ }Q_{1 }$(2)&7908.622c&2.43&        &      &\\
\cline{1-4}
$\rm ^{Q}R_{12}$(1)&7908.959c&4.25&2.76N(1)& 53.3 &polluted\\
$\rm ^{ }Q_{1 }$(1)&7908.966c&2.02&        &      &\\
\cline{1-4}
$\rm ^{P}Q_{12}$(2)&7913.692c&1.60&0.86N(2)& 19.1 &\\
$\rm ^{ }P_{1 }$(2)&7913.704c&0.36&        &      &\\
\cline{1-4}
$\rm ^{P}Q_{12}$(3)&7915.714c&1.70&1.08N(3)& 18.5 &\\
$\rm ^{ }P_{1 }$(3)&7915.729c&0.61&        &      &\\
\cline{1-4}
$\rm ^{O}P_{12}$(3)&7920.804c&0.25&0.11N(3)& nd   &\\
\hline
\hline
\multicolumn{5}{l}{no: not observed thus not available} \\
\end{tabular}}
\end{table*}

\begin{table*} % Table~7, not in article
\caption{CN Red System (3,0) with  $f_{(3,0)}=3.34\times 10^{-4}$.
Derived column densities are given in Table~4.
Table will not appear in journal.}
\centerline{
\begin{tabular}{rllllll}
\hline
\hline
B($N''$)&$\lambda_{\rm rest}$[\AA]&$f_{N'J',N''J''}\times 10^{4}$ &
$f_{\rm eff}\times 10^{4}N(N'')$ [cm$^{-2}$]
&$W_{\lambda}$[m\AA]&Remark\\
\hline   
$^{S}R_{21}$(4)&6916.868l&0.282&0.157N(4)&no          &\\
\cline{1-4}
$^{S}R_{21}$(3)&6918.581l&0.310&0.177N(3)&no          &\\
\cline{1-4}
$^{S}R_{21}$(2)&6920.480l&0.343&0.206N(2)&no          &\\
\cline{1-4}
$^{S}R_{21}$(1)&6922.568l&0.387&0.258N(1)&no          &\\
\cline{1-4}
$^{S}R_{21}$(0)&6924.827l&0.461&0.461N(0)&$10.7\pm1.0$&\\
\cline{1-4}
$    R_{2 }$(4)&6925.811c&0.627&         &            &\\
$^{R}Q_{21}$(4)&6925.821l&0.612&0.619N(4)&$ 1.9\pm0.5$&tentative \\
\cline{1-4}
$    R_{2 }$(3)&6925.893c&0.619&         &            &\\
$^{R}Q_{21}$(3)&6925.909l&0.663&0.644N(3)&$ 6.5\pm0.5$&\\
\cline{1-4}
$    R_{2 }$(2)&6926.194c&0.622&         &            &\\
$^{R}Q_{21}$(2)&6926.178l&0.725&0.684N(2)&$16.3\pm0.5$&\\
\cline{1-4}
$    R_{2 }$(1)&6926.656c&0.658&         &            &\\
$^{R}Q_{21}$(1)&6926.650l&0.811&0.760N(1)&$23.0\pm0.5$&\\
\cline{1-4}
$^{R}Q_{21}$(0)&6927.304l&1.110&1.110N(0)&$48.9\pm5.0$&polluted\\
\cline{1-4}
$    Q_{2 }$(1)&6929.062l&1.110&0.555N(1)&$18.8\pm0.5$&\\
$^{Q}P_{21}$(1)&6929.067c&0.278&         &            &\\
\cline{1-4}
$    Q_{2 }$(2)&6930.272l&0.969&0.576N(2)&$14.0\pm0.5$&\\
$^{Q}P_{21}$(2)&6930.302c&0.314&         &            &\\
\cline{1-4}
$    Q_{2 }$(3)&6931.673l&0.992&0.605N(3)&$ 6.0\pm1.0$&line too weak \\
$^{Q}P_{21}$(3)&6931.663c&0.315&         &            &\\
\cline{1-4}
$    P_{2 }$(2)&6932.715l&0.278&0.111N(2)&$ 6.0\pm0.5$&\\
\cline{1-4}
$    Q_{2 }$(4)&6933.179l&1.030&0.627N(4)&$\leq 1.5$  &\\
$^{Q}P_{21}$(4)&6933.206c&0.305&         &            &\\
\cline{1-4}
$    P_{2 }$(3)&6935.720l&0.353&0.151N(3)&$\leq 1.5$  &\\
\cline{1-4}
$    P_{2 }$(4)&6938.910l&0.399&0.177N(4)&$\leq 1.5$  &\\
\cline{1-4}
$    R_{1 }$(4)&6945.273l&0.811&0.451N(4)&$\leq 1.5$  &\\
\cline{1-4}
$    R_{1 }$(3)&6946.479l&0.850&0.486N(3)&$ 5.9\pm1.0$&\\
\cline{1-4}
$    R_{1 }$(2)&6947.962l&0.929&0.557N(2)&$13.0\pm0.5$&polluted\\
\cline{1-4}
$    R_{1 }$(1)&6949.763l&1.115&0.743N(1)&$24.0\pm1.0$&\\
\cline{1-4}
$    R_{1 }$(0)&6951.816l&1.766&1.766N(0)&$24.0\pm5.0$&\\
\cline{1-4}
$^{Q}R_{12}$(2)&6953.415c&0.881&         &            &\\
$    Q_{1 }$(2)&6953.461l&0.898&         &$24.1\pm0.5$&\\
$^{Q}R_{12}$(3)&6953.459c&0.655&         &            &\\
$    Q_{1 }$(3)&6953.461l&0.980&         &            &\\
\cline{1-4}
$^{Q}R_{12}$(1)&6953.643c&1.568&1.021N(1)&$25.7\pm0.5$&\\
$    Q_{1 }$(1)&6953.643l&0.747&         &            &\\
\cline{1-4}
$^{Q}R_{12}$(4)&6953.776c&0.532&         &            &\\
$    Q_{1 }$(4)&6953.787l&1.040&0.814N(4)&$ 2.3\pm0.5$&tentative \\
\cline{1-4}
$^{P}Q_{12}$(2)&6957.302c&0.593&0.316N(2)&$ 8.4\pm0.5$&\\
$    P_{1 }$(2)&6957.301l&0.132&         &            &\\
\cline{1-4}
$^{P}Q_{12}$(3)&6958.905c&0.631&0.397N(3)&$ 4.6\pm0.5$&\\
$    P_{1 }$(3)&6958.907l&0.222&         &            &\\
\cline{1-4}
$^{P}Q_{12}$(4)&6960.782c&0.612&0.433N(4)&$\leq 1.5$  &\\
$    P_{1 }$(4)&6960.805l&0.289&         &            &\\
\cline{1-4}
$^{O}P_{12}$(3)&6962.795l&0.092&0.039N(3)&$\leq 1.5$  &\\
\cline{1-4}
$^{O}P_{12}$(4)&6966.212l&0.138&0.061N(4)&$\leq 1.5$  &\\
\hline
\hline
\multicolumn{5}{l}{no: not observed thus not available} \\
\end{tabular}}
\end{table*}

\end{document}